\documentclass[acmsmall]{acmart}

\AtBeginDocument{%
  \providecommand\BibTeX{{%
    \normalfont B\kern-0.5em{\scshape i\kern-0.25em b}\kern-0.8em\TeX}}}

\setcopyright{acmlicensed}
\acmJournal{PACMHCI}
\acmYear{2020} \acmVolume{4} \acmNumber{CSCW2} \acmArticle{149} \acmMonth{10} \acmPrice{15.00}\acmDOI{10.1145/3415220}

\usepackage{balance}       
\usepackage{graphics}      
\usepackage[T1]{fontenc}   
\usepackage{booktabs}
\usepackage{textcomp}
\usepackage{multirow}
\usepackage{graphicx}
\usepackage{tabularx,ragged2e}
\usepackage{wrapfig}
\usepackage{subfig}
\usepackage{arydshln}
\usepackage{caption}
\usepackage{varwidth}
\usepackage{color}
\usepackage[T1]{fontenc}
\usepackage{placeins}

\usepackage{microtype}        
\usepackage{ccicons}          
\usepackage[utf8]{inputenc} 

\usepackage{todonotes}

\RequirePackage[loading]{tracefnt}

\definecolor{linkColor}{RGB}{6,125,233}

\begin{document}

\title{Vision Skills Needed to Answer Visual Questions}

\author{Xiaoyu Zeng}
\email{edith.xiaoyu.zeng@utexas.edu}
\affiliation{%
  \institution{The University of Texas at Austin}
  \streetaddress{School of Information}
  \city{Austin}
  \state{Texas}
  \postcode{78701}
  \country{USA}
}

\author{Yanan Wang}
\affiliation{%
  \institution{University of Wisconsin-Madison }
  \streetaddress{Department of Computer Science}
  \city{Madison}
  \state{Wisconsin}
  \postcode{53706}
  \country{USA}}
\email{yanan@cs.wisc.edu}

\author{Tai-Yin Chiu}
\affiliation{%
  \institution{The University of Texas at Austin}
  \streetaddress{}
  \city{Austin}
  \state{Texas}
  \postcode{78701}
  \country{USA}
}
\email{chiu.taiyin@utexas.edu}

\author{Nilavra Bhattacharya}
\affiliation{%
  \institution{The University of Texas at Austin}
  \streetaddress{School of Information}
  \city{Austin}
  \state{Texas}
  \postcode{78701}
  \country{USA}
}
\email{nilavra@ieee.org}

\author{Danna Gurari}
\affiliation{%
  \institution{The University of Texas at Austin}
  \streetaddress{School of Information}
  \city{Austin}
  \state{Texas}
  \postcode{78701}
  \country{USA}
}
\email{danna.gurari@ischool.utexas.edu}

\renewcommand{\shortauthors}{Zeng et al.}

\begin{abstract}
The task of answering questions about images has garnered attention as a practical service for assisting populations with visual impairments as well as a visual Turing test for the artificial intelligence community.  Our first aim is to identify the common vision skills needed for both scenarios.  To do so, we analyze the need for four vision skills---object recognition, text recognition, color recognition, and counting---on over 27,000 visual questions from two datasets representing both scenarios.  We next quantify the difficulty of these skills for both humans and computers on both datasets.  Finally, we propose a novel task of predicting what vision skills are needed to answer a question about an image.  Our results reveal (mis)matches between aims of real users of such services and the focus of the AI community.  We conclude with a discussion about future directions for addressing the visual question answering task.
\end{abstract}

\keywords{Visual Question Answering; Computer Vision; Accessibility}

\ccsdesc[500]{Information systems~Crowdsourcing}
\ccsdesc[500]{Computing methodologies~Computer vision}

\maketitle

\section {Introduction}
In visual question answering (VQA), the task is to answer a question about an image.  Technologies that perform this task emerged around 2010, when people with visual impairments were first empowered to rely on personal mobile devices to solicit information about their visual surroundings from remote crowds~\cite{bigham2010VizWiznearlyrealtime}.  A number of new VQA services have emerged since~\cite{BeSpecular,HomeAiraAira,bemyeyes}.  Such VQA technologies serve as game changers for this community.  Rather than having to expend social capital to request information from friends or family, users can rapidly make decisions about what to eat, wear, purchase, and more. Users' access to such visual assistants empower them to meet their information seeking goals more independently.

Drawing inspiration from human-based VQA services, the artificial intelligence (AI) community took on the challenge of developing algorithms that can automatically answer visual questions.  This trend took off in 2015, with the release of a community-wide challenge around a large-scale VQA dataset paired with an annual workshop since 2016 to chart and celebrate the community's progress~\cite{antol2015VqaVisualquestion}.  

Despite the widespread goal to assist people with visual impairments, few publications reveal how this population uses such technologies and the relevance of AI community's efforts in addressing real users' interests.  We aim to gain a more nuanced understanding by examining this through the lens of the \textit{visual skills} needed to solve the visual question answering task. Skills reveal the (un)common uses and expected (in)abilities of VQA services.

\begin{figure*}[t!]
    \centering
    \includegraphics[width=\textwidth]{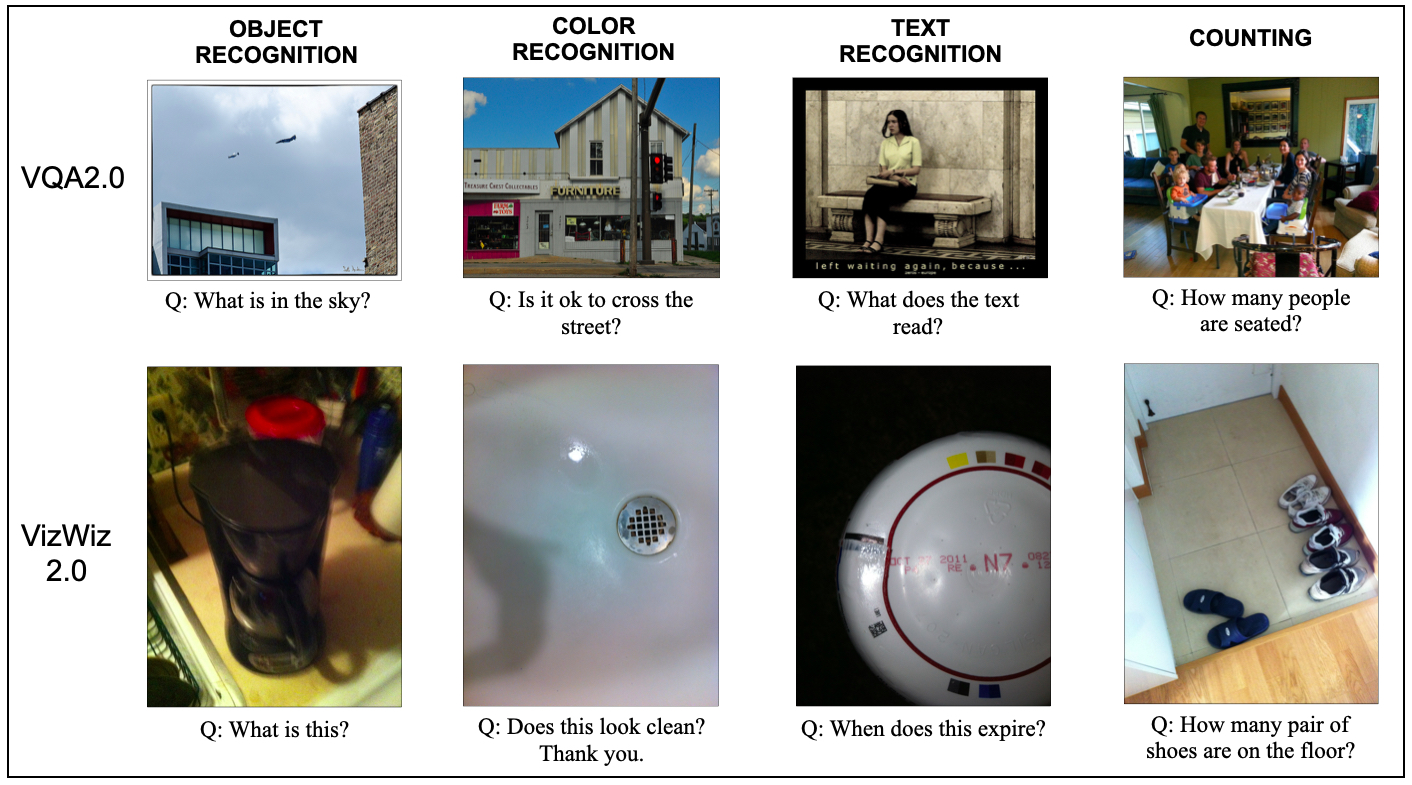}
    \caption{Examples of visual questions used for algorithm development in the AI community (top row; VQA2.0~\cite{making_the_v_in_vqa_matter}) and visual questions asked by real users of VQA services (bottom row; VizWiz~\cite{gurari2018VizWizGrandChallenge}).  Shown are examples for each of the four vision skills analyzed in this paper.}
    \label{fig:vq_examples}
\end{figure*}

Our first contribution is to analyze and compare the vision skills needed for two VQA datasets that reflect two mainstream use cases of VQA: assisting real users and facilitating the AI community's research.  To do so, we employed crowdsourcing to label which skills are needed to answer a visual question for over 27,000 visual questions from two datasets.  We focused on the following four vision-based skills that are inspired by prior work~\cite{antol2015VqaVisualquestion,brady2013Visualchallengeseveryday} and pilot studies: object recognition, text recognition, color recognition, and counting (exemplified in Figure~\ref{fig:vq_examples}).  We examined the following questions: (1) What are the skills that real users of VQA services most commonly need assistance with?, (2) What are the skills that are most prominent in AI datasets for training and evaluating VQA algorithms?, and (3) To what extent do the skills in each scenario overlap?  

Our next contribution is to gauge what makes it easy versus difficult to answer visual questions.  We benchmarked the performance of humans and algorithms in providing answers for visual questions that involve each skill as an indicator of each skill's difficulty.  Our analysis has immediate implications for what to improve in VQA services.  For human-powered answering services, it reveals relevant skill training that is needed.  For algorithm development, our findings highlight to what extent algorithms need to be improved to tackle each skill.  Our findings can also be valuable in setting users' expectations regarding the limits of publicly-available automated visual assistance technology today.

Finally, our work is inspired by the longer-term goal of improving the design of VQA frameworks.  We propose a novel task of predicting what skills are needed to answer a visual question, and an algorithm for this purpose.  We envision this task can serve as a valuable foundation for a variety of improved VQA services.  One approach is to assign visual questions to human workers based on their preferences and expertise for each skill.  Another approach is to assign visual questions to various applications and vision APIs that are optimized for each specific skill.  A third approach is to combine the previous two approaches and dynamically connect each visual question with the appropriate algorithm if one is suitable or appropriate human otherwise.  We will publicly share our new VQA-Skills dataset and code to encourage these and other future directions of work (\texttt{https://vizwiz.org}).\footnote{The shared content consists of the labels indicating which skills are needed to answer each visual question (i.e., dataset) and the machine learning algorithm that predicts what vision-based skills are needed to answer a visual question (i.e., code).}

Altogether, our work offers valuable insights to multiple stakeholders — creators of computer vision datasets, designers of VQA systems, and users of VQA systems.  It is broadly-known that computer vision datasets can embed flawed assumptions that lead to perpetuating and amplifying biases in technology, including for VQA~\cite{chao2018Crossdatasetadaptationvisual,das2019Datasetbiascase}.  Yet, the types of biases can be unknown to those developing the AI.  That is because recognizing these biases can require understanding real users' interests to assess which interests are not reflected in datasets used by the AI community.  Our work raises awareness about a practically important challenge of narrowing the divide between what we should support socially and what we can expect from technology today.  
\section {Related Work}
\paragraph{Visual Question Answering.}
Visual question answering has been examined both as a practical service to assist populations with visual impairments~\cite{bigham2010VizWiznearlyrealtime} as well as a Visual Turing Test~\cite{visual-turing-test,malinowski2014multiworldapproachquestion} to measure progress in the AI community in learning how to emulate people's vision and language perception capabilities~\cite{malinowski2014multiworldapproachquestion}.  The interest in both of these purposes has been growing, as exemplified by the increasing number of visual assistance applications~\cite{BeSpecular,HomeAiraAira,bemyeyes} alongside the AI community's growing number of public VQA benchmark datasets~\cite{kafle2017visual,wu2017survey}, and VQA algorithms~\cite{kafle2017visual,wu2017survey}.  Yet, the open-ended nature of the VQA task has led some researchers to try to demystify what this task entails.  For example, Brady et al.~\cite{brady2013Visualchallengeseveryday} analyzed 1,000 visual questions from real users of VQA services and classified them as primarily spanning requests to identify objects, describe objects, and read text.  The AI community have embraced the division of such visual questions based on the types of answers they elicit: ``yes/no'', ``number''  or ``other''~\cite{antol2015VqaVisualquestion}.  Complementing prior work, we contribute knowledge about the vision-based skills that are needed to answer visual questions asked by real users of VQA services and in the most popular VQA dataset.  In doing so, our findings offer a more nuanced understanding of the real needs of VQA users, the focus of the AI community, and how well the two align.

\paragraph{Understanding Dataset Bias.}  
More generally, our work relates to the plethora of prior work that exposes biased datasets that lead to biased algorithms.  For example, bias has been analyzed for commercial video clips~\cite{jang2019QuantificationGenderRepresentation}, automated facial analysis technologies~\cite{scheuerman2019HowComputersSee}, algorithms that predict whether a convicted criminal will re-offend~\cite{vanberkel2019CrowdsourcingPerceptionsFair}, loan lending algorithms~\cite{passi2018TrustDataScience}, and LabInTheWild studies~\cite{august2018FramingEffectsChoice}.
 More similar to our work, it has been shown that it is possible, for a given visual question, to identify which dataset it belongs to from a known collection of VQA datasets~\cite{chao2018Crossdatasetadaptationvisual,das2019Datasetbiascase}.  Similar to prior work, our findings expose biases in datasets, and so underscore important next steps to take towards mitigating them.  Unlike prior work, we identify the skills real users seek that are nevertheless ill-represented in the mainstream VQA dataset used to train machine learning algorithms.

\paragraph{Analyzing Difficulty in Answering Visual Questions}
Several research groups have discussed ideas related to establishing what are easier versus more difficult visual questions.  Some researchers discuss this from the lens of the challenges faced by \emph{humans} in answering visual questions.  For example, Antol et al.~\cite{antol2015VqaVisualquestion} asked crowds to identify what age (and so level of visual training) is required in order to answer visual questions, and reported that the levels for 10,000 visual questions were as follows: 15.3\% for toddler, 39.7\% for younger child, 28.4\% for older child, 11.2\% for teenager, 5.5\% for adult.  Another group quantified how often different humans return the same answer to a visual question~\cite{gurari2017CrowdVergePredictingIf}, and reported that this arises for roughly half of the $\sim$450,000 visual questions analyzed.  And another group examined why different answers to a visual question arise~\cite{bhattacharya2019does}.  Other researchers discuss visual question difficulty from the lens of the challenges faced by \emph{computers} in answering visual questions.  For example, Yeh et al.~\cite{yeh2008photo} measured the difficulty to find a similar matching visual question as a guide to provide a suitable answer.  Goyal et al. reported upon an algorithm's ability to consistently return the same answer when slightly altering the question~\cite{goyal2016towards}.  Our work complements prior work in that we offer analysis that highlights the difficulty of different vision-based skills for humans and computers.  More generally, our work contributes to the literature that seeks to characterize the difficulty of various tasks for computers and humans~\cite{zhang2019Dissonancehumanmachine}.

\paragraph{Accessibility}
Many applications assist people with visual impairments to more independently learn about the visual world around them.  Often such solutions rely on having humans-in-the-loop (i.e., partially or fully rely on human intelligence to describe visual information)~\cite{bigham2010VizWiznearlyrealtime, bigham2010locateIt,bemyeyes,guinness2018CaptionCrawlerEnabling,guo2016vizlens,taptapsee,sudol2010looktel,zhong2015RegionspeakQuickcomprehensive} in order to consistently offer accurate results.  However, the potential high cost, high latency, and privacy concerns can be significant drawbacks.  Accordingly, many computer vision researchers are developing machines to replicate the skills of the human vision system in order to assist people who are blind~\cite{kacorri2017Peoplevisualimpairment,salisbury2017toward}. Current automated solutions that are used in practice (e.g., BlindSight \cite{blindsight}, Seeing AI \cite{seeingai}, Envision \cite{envisionai}, Orcam~\cite{Orcam}) offer algorithms that are targeted towards specific skills; for instance, to recognize an object category, read text, etc.  That is because existing VQA algorithms are too inaccurate to be useful in practice at present.  Towards bridging the gap between VQA and existing assistive tools used in practice, we uncover current limitations of mainstream VQA datasets and methods.  We also introduce a novel problem and implementation of a skill prediction system as a valuable intermediary step towards enabling users to ask open-ended questions in order to decide which of the set of predetermined computer vision tools to deploy.  Our experiments demonstrate the feasibility of this prediction task (i.e., by showing that trained models can considerably outperform random guessing).  More generally, our work fits into the broad collection of work that seeks to better understand and meet the needs of people with visual impairments~\cite{baldwin2019DesignPublicSquare,das2019Itdoesnwin,shi2019AccessibleVideoCalling,stangl2018BrowseWithMeOnlineClothes,gurari2018PredictingForegroundObject,gurari2019VizWizPrivDatasetRecognizing,gurari2020captioning,chiu2020assessing,stangl2020person}.
\section{VQA Datasets: Needed Vision Skills}
Our first aim is to better understand what vision-based skills are needed to answer visual questions asked by real users as well as those commonly used to develop VQA algorithms.  For both settings, we investigate (1) how often is each skill needed?, (2) how common skills are standalone versus occurring with other skills?, and (3) what types of questions are commonly asked for each skill type?  Our findings offer new insights about real users' information wants, skills  the AI community is teaching VQA algorithms, and how well the two interests align.

\subsection{Labeling VQA Datasets}
We now describe our process for assigning skills to visual questions as the foundation for our analysis. 

\paragraph{Source of VQAs}
We chose to analyze 27,263 visual questions from two modern datasets: VizWiz~\cite{gurari2018VizWizGrandChallenge} and VQA2.0~\cite{making_the_v_in_vqa_matter}.  While these datasets are similar in that each visual question consists of an image, a question about the image, and 10 crowdsourced answers, they are different in how the VQAs were generated.  Specifically, the datasets were generated as follows:

\begin{itemize}
    \item \textit{VizWiz}~\cite{gurari2018VizWizGrandChallenge}: This dataset is built upon a collection of visual questions asked by real users of a mobile phone application called VizWiz~\cite{bigham2010VizWiznearlyrealtime}.  With this application, a user takes a picture, records a spoken question, and then submits them to receive answers from remote human respondents.  This application was designed to empower people who are blind to quickly learn about their physical surroundings, and the prevailing assumption is that all the visual questions come from that population.  Of note, we excluded visual questions that are unanswerable since we cannot know what skills are needed for these.  We identified unanswerable questions as those visual questions which the crowd workers supplying the answers flagged as having images that are ``too poor in quality to answer the question (i.e., all white, all black, or too blurry)'' or for which ``the question cannot be answered from the image.'' In total, we used the 22,204 visual questions that are answerable.
    
    \item \textit{VQA2.0}~\cite{making_the_v_in_vqa_matter}: This dataset was inspired in part to support the interests of people who are blind, but was more generally designed to serve as a large-scale dataset for the AI community.  It is the most widely used dataset for AI visual question answering research, having attracted over 120 teams since 2016 who have competed on this dataset challenge.  It was built upon a curated collection of images in the MS COCO \cite{MSCOCO} dataset which were originally collected to support the object recognition task. MS COCO was developed by scraping Google, Bing, and Flickr for images that show complex scenes that contain at least one of 80 common object categories that would be easily recognizable by a four-year-old.  Subsequently, crowd workers were recruited to generate questions about the images that would ``stump [a] smart robot'' \cite{antol2015VqaVisualquestion}.  We sampled a total of 5,034 visual questions for our analysis.  To capture the diversity in this dataset, we performed stratified sampling with respect to different types of answers and types of questions.\footnote{Answer types were included with the dataset, with each visual question assigned to one of the following answer types: ``yes/no,'' ``number,'' or ``other.''  The type assigned to each visual question is the most popular from 10 labels that are assigned to each of its 10 answers.  834 visual questions were randomly sampled for each answer type.  Question type refers to what kind of question is asked (``how many...'', ``what is...'', etc.) based on up to the first six words of the question.}
\end{itemize}

In summary, these datasets reflect very different circumstances.  While VizWiz represents the interests of real users of VQA services, VQA2.0 represents the focus of the AI community.  Additionally, VizWiz represents humans sharing their own interests with remote humans voluntarily, while VQA2.0 represents paid humans testing a computer's intelligence.  Yet an important motivation for the VQA2.0 dataset is to empower the AI community to focus on technical challenges that are ``... applicable to scenarios encountered when visually-impaired users... actively elicit visual information.''  Our comparison of these two datasets simultaneously can offer a reality check about the extent to which today's mainstream dataset for the AI community can meet this aim and reveal opportunities to design future large-scale VQA datasets to meet this aim.  

\paragraph{Skill Categories}
To our knowledge, no prior work has analyzed what vision skills are needed to answer visual questions.  Consequently, we developed a taxonomy for this purpose.  We chose to focus on high-level, human-interpretable vision skills (as opposed to low-level skills such as finding edges or shapes).  

Our taxonomy took inspiration from prior work and was iteratively refined based on pilot studies.  We initially chose three categories based on a classification posed by Brady et al.~\cite{brady2013Visualchallengeseveryday} regarding the types of questions people who are blind asked about 1,000 images: to identify objects, get object descriptions (majority of which were for the color), and read text.  We ran a pilot study with 1,000 VQAs randomly sampled from both datasets using a preliminary version of the crowdsourcing system (discussed in the next paragraph) to assign the categories of ``object recognition,'' ``text recognition,'' ``color recognition'' or ``other'' paired with a free-text entry box.  Based on crowd workers' feedback, we introduced a fourth category that was commonly observed for the ``other'' category: counting.  We then conducted a second pilot study with 400 randomly sampled VQAs from both datasets to verify the robustness of this taxonomy.  This process culminated in our final taxonomy, which includes the following four skills: object recognition, text recognition, color recognition, and counting.  

\begin{figure}[t!]
    \centering
    \includegraphics[width=0.82\columnwidth]{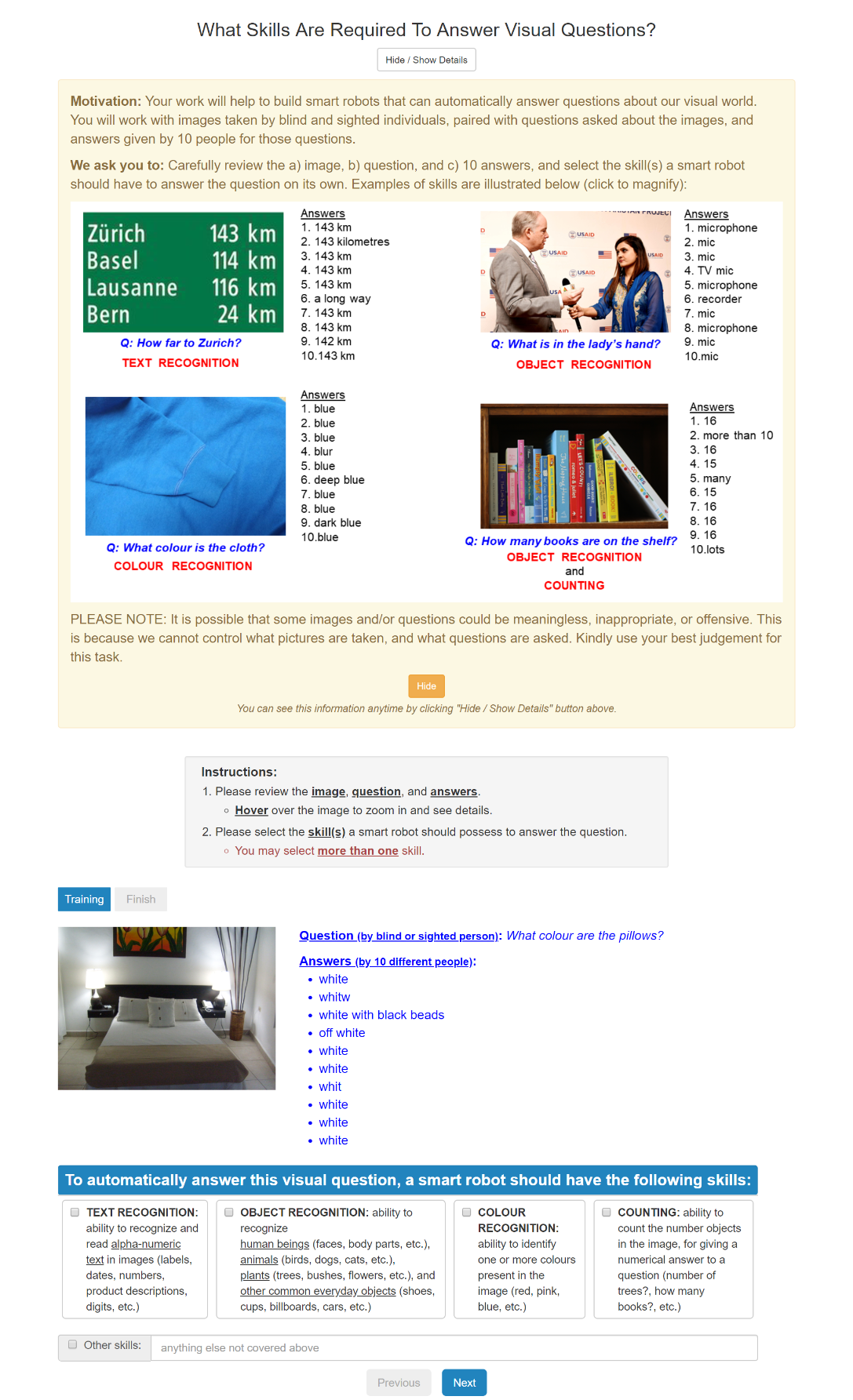}
    \caption{Screen shot showing the user interface for our crowdsourcing task.}
    \label{fig:interface}
\end{figure}

\paragraph{Assigning Skills to Visual Questions}
We recruited crowd workers to identify what skills are needed to answer each visual question.  An example of the labeling task we asked them to complete is in Figure~\ref{fig:interface}.  The user was shown a VQA (image, question, and ten answers) and asked to select all skills that are needed to provide the answers.  To help users understand how to complete the task, we included step-by-step instructions and four labeled examples.  
To explicitly convey that multiple reasons could be selected when more than one reason was applicable, we included both an example that illustrates this scenario and wrote ``You may select more than one skill.'' We also provided a comments box, as a free-entry text-box, so that users could leave feedback if they wanted.  

We employed crowd workers from Amazon Mechanical Turk to perform our task.  For quality control, we only accepted workers who previously completed more than 500 jobs, had at least a $95\%$ approval rating, and were from the US (to try to ensure English proficiency).  We also collected redundant labels, asking five independent crowd workers to label each visual question, and then used the majority vote decision per skill to decide which skills are relevant.  In total, 475 workers contributed 718.67 person-hours to complete all the 136,315 crowdsourced tasks (i.e., 5 people $\times$ $27,263$ VQAs).  Average pay was $\$7.53$ per hour.  
\subsection{Skill Prevalence}
We first tallied how often each skill is needed to answer visual questions asked by real users and visual questions used for training VQA algorithms.   

Nearly all visual questions from our datasets require at least one of the four skills; i.e., $99.13\%$ of visual questions in VizWiz and $99.17\%$ of visual questions in VQA2.0 have at least one skill label chosen by crowd workers.  The ``other'' category accounts for a negligible amount of visual questions (2 questions in VizWiz and 5 questions in VQA2.0). These findings suggest that our skill taxonomy provides sufficient coverage for the visual questions.

For visual questions from real users (i.e., VizWiz dataset), the most common skill is object recognition (93.25\%), followed by text recognition (45.33\%), color recognition (22.06\%), and counting (1.49\%).  Our findings differ considerably from those reported by the related work of Brady et al.~\cite{brady2013Visualchallengeseveryday}, which discussed the prevalence of different types of questions asked: 41\% sought object recognition (``Identification''), 17\% text recognition (``Reading''), and 11.5\% color recognition (subset of ``description" questions about clothing and object color).  Our percentages are roughly double that reported in prior work.  We attribute our different findings largely to the fact that our work examines a somewhat distinct task from that examined by Brady et al.~\cite{brady2013Visualchallengeseveryday}.  While prior work focused on the final task alone, we instead considered all skills needed throughout the process which includes the intermediary steps needed to perform the final task.   Moreover, while prior work permitted only one category per visual question, our work permitted multiple categories per visual question.  A further distinction is that our analysis is based on a considerably larger amount of data; that is, a 22-fold increase in the number of visual questions.

For visual questions commonly employed to train VQA algorithms (i.e., VQA2.0 dataset), the most common skill is object recognition (97.91\%), followed by color recognition (16.7\%), counting (16.56\%), and text recognition (5.72\%).  The most similar analysis to ours is by Antol et al.~\cite{antol2015VqaVisualquestion}, which reported the percentage of visual questions that involve counting.  Our findings complement this work in that we found a similar percentage to what they report (i.e., 12.31\%~\cite{antol2015VqaVisualquestion}).  Our work also goes further by offering greater insight about the dataset for more factors than what have been previously analyzed.

We next compare our findings across both settings (i.e., VizWiz and VQA2.0).  We observe similar trends for two of the skills.  Object recognition is consistently the most popular skill, with over $93\%$ of visual questions in both datasets requiring it.  This suggests that having the ability to recognize objects is typically a prerequisite skill.  Color recognition also is a popular skill with similar frequencies in both datasets ($16.7\%$ in VQA2.0 and $22.06\%$ in VizWiz),  In contrast, we observe different trends for the two other skills. We noticed that text recognition skill is $7.9$ times \emph{more likely} in VizWiz than in VQA2.0. The result of our Kruskal-Wallis H-test confirms that the difference between text recognition's prevalence in the two datasets is statistically significant ($F(1, \,27,258) = 2735.41, p < .001$ and Welch's $t = 84.81, p < .001, N_{\text{VizWiz}} = 22,226, N_{\text{VQA2.0}} = 5,034$). Likewise, counting is $11.1$ times \emph{less likely} in VizWiz than in VQA2.0, with statistically significant Kruskal-Wallis H-test results ($F(1, \,27,258) = 2280.84, p < .001$ and Welch's $t = -28.43, p < .001, N_{\text{VizWiz}} = 22,226, N_{\text{VQA2.0}} = 5,034$). These differences highlight a considerable mismatch in the skills that the AI community is focusing on teaching their algorithms and the skills that real users of VQA services would want from AI algorithms.
 
\subsection{Skill Co-occurrence}
\begin{figure}[t!]
\centering
\includegraphics[width=0.6\textwidth]{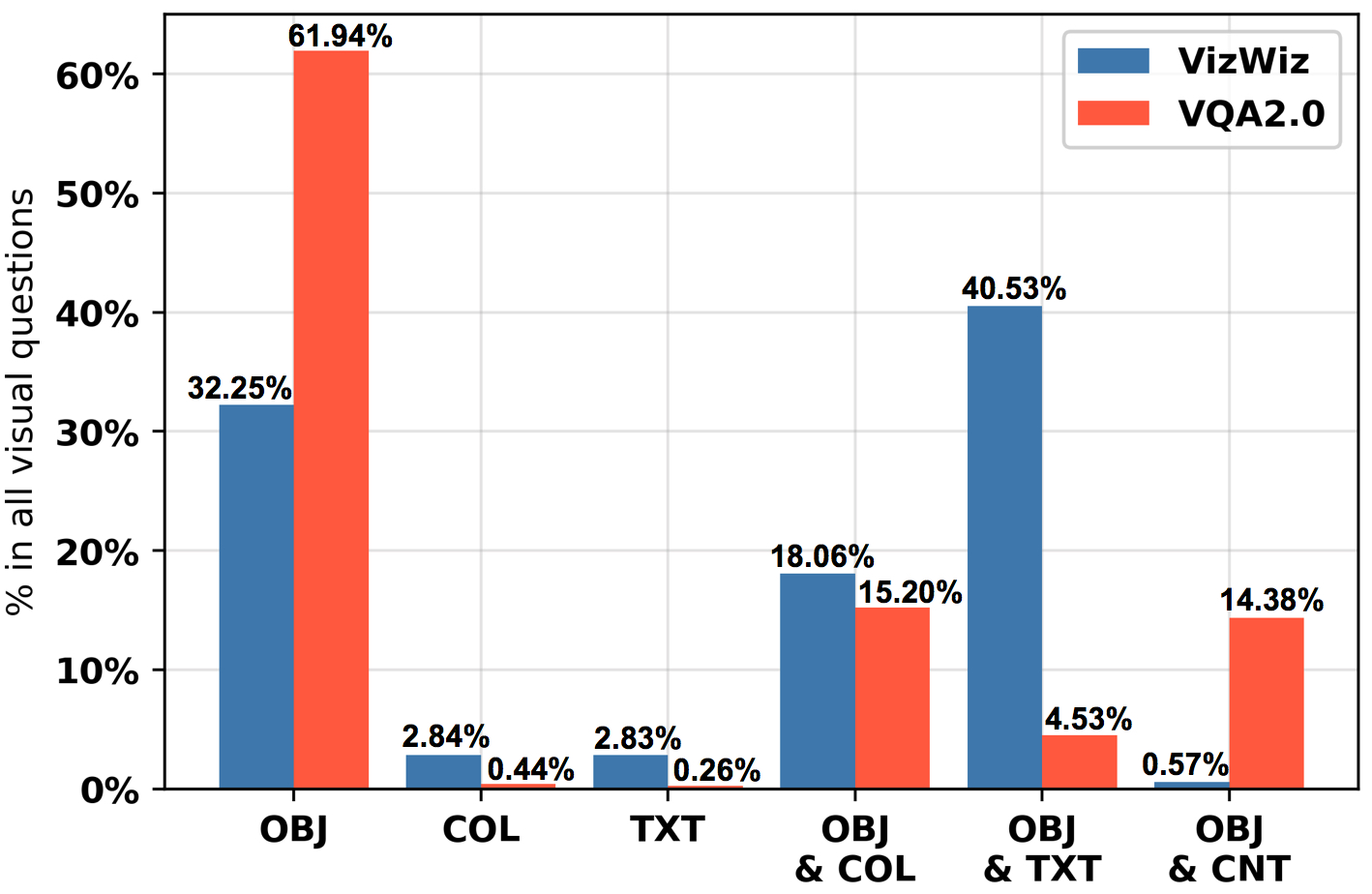}
\caption{Shown is the frequency of all skill combinations, excluding those that account for less than 1\% of visual questions in both datasets.  Combinations not shown in the figure are as follows: TXT \& COL (0.91\% of VizWiz; 0.14\% of VQA), TXT \& CNT (0.84\%; 0.77\%), TXT \& COL (0.94\%; 0.14\%), COL \& CNT (0.21\%; 0.91\%), OBJ \& TXT \& COL (0.91\%; 0.14\%), OBJ \& COL \& CNT (0.20\%; 0.89\%), TXT \& COL \& CNT (0\% in both datasets), and OBJ \& TXT \& COL \& CNT (less than $0.01\%$ in both datasets).}
\label{fig:skill_comb_dist}
\end{figure}

We next quantified the extent to which the skills co-occur versus arise alone.  To do so, we calculated frequencies of all possible skill combinations for both datasets.  Results are shown in Figure~\ref{fig:skill_comb_dist}.
 
For visual questions from real users (i.e., VizWiz dataset), we observe that most commonly object recognition and text recognition are needed jointly (40.53\%), followed by object recognition alone (32.25\%), and object recognition with color recognition (18.06\%).  All other possible skill combinations arise for less than 3\% of visual questions each. This indicates that object recognition algorithms are necessary but often insufficient alone to answer most real-world visual questions.  The percentage of questions that require object recognition with just one other skill is approximately $59\%$ in VizWiz.\footnote{Calculated from Figure~\ref{fig:skill_comb_dist} as the sum of $18.06\%$ for object and color recognition, $40.53\%$ for object and text recognition, and $0.57\%$ for object and counting.}  This finding is exciting because it suggests that, even though object recognition alone is not enough to solve most real visual questions, over half of those questions can be solved with algorithms specifically designed for just one additional skill.  Questions that require three skills (2\%) and all four skills (less than $0.01\%$) are very rare.

For visual questions used to train VQA algorithms (i.e., VQA2.0 dataset), we observe that most commonly object recognition alone is needed (61.94\%), followed by object recognition with color recognition (15.2\%), and object recognition with counting (14.38\%).  All other possible skill combinations arise for less than 5\% of visual questions each.  This reveals that object recognition algorithms are often sufficient to answer most visual questions used to develop VQA algorithms.  The percentage of questions that require more than one skill is approximately $34\%$ for two skills, $1.83\%$ for three skills, and less than $0.01\%$ for all four skills.

Our key observation when comparing our findings across both settings (i.e., VizWiz and VQA2.0) is that real users' visual questions tend to be more complex than those in the mainstream VQA dataset for the AI community.  Specifically, they require more than one skill almost twice as often (i.e., $62\%$ for VizWiz versus $37\%$ for VQA2.0).  Skill complexity appears to be comparable across both datasets only for color recognition, as evidenced by its similar frequency when alone and paired with object recognition.  In other words, the mainstream VQA dataset in the AI community may be a good data source for representing some of real users' interests, centering around color recognition.

Another interesting observation is that most of the skills tend to co-occur with the other skills.  This is especially interesting counting, because algorithm developers often report VQA algorithm performance based on whether the answer type is a number and so deemed to be a ``counting'' question (e.g., ~\cite{antol2015VqaVisualquestion,gurari2018VizWizGrandChallenge,making_the_v_in_vqa_matter}).  Our findings underscore that such evaluations actually measure algorithms' abilities to perform a combination of skills.  In particular, counting often involves first identifying the object and possibly its attributes.  For example, questions such as ``How many pink lines are on this test strip?'' (VizWiz) and ``How many propellers are there?'' (VQA2.0), assume that the object has already been identified; ``How much money is it?'' (VizWiz) and ``How many colors can be seen in this picture?'' (VQA2.0) require the answerer to tally numerous banknotes and colors.  Without color, text, or object recognition skills, less than $0.03\%$ of VizWiz and $0.5\%$ of VQA2.0 need counting.  
 
\subsection{Questions Asked for Each Skill}
\begin{figure*}[t!]
    \centering
    \includegraphics[width=\textwidth]{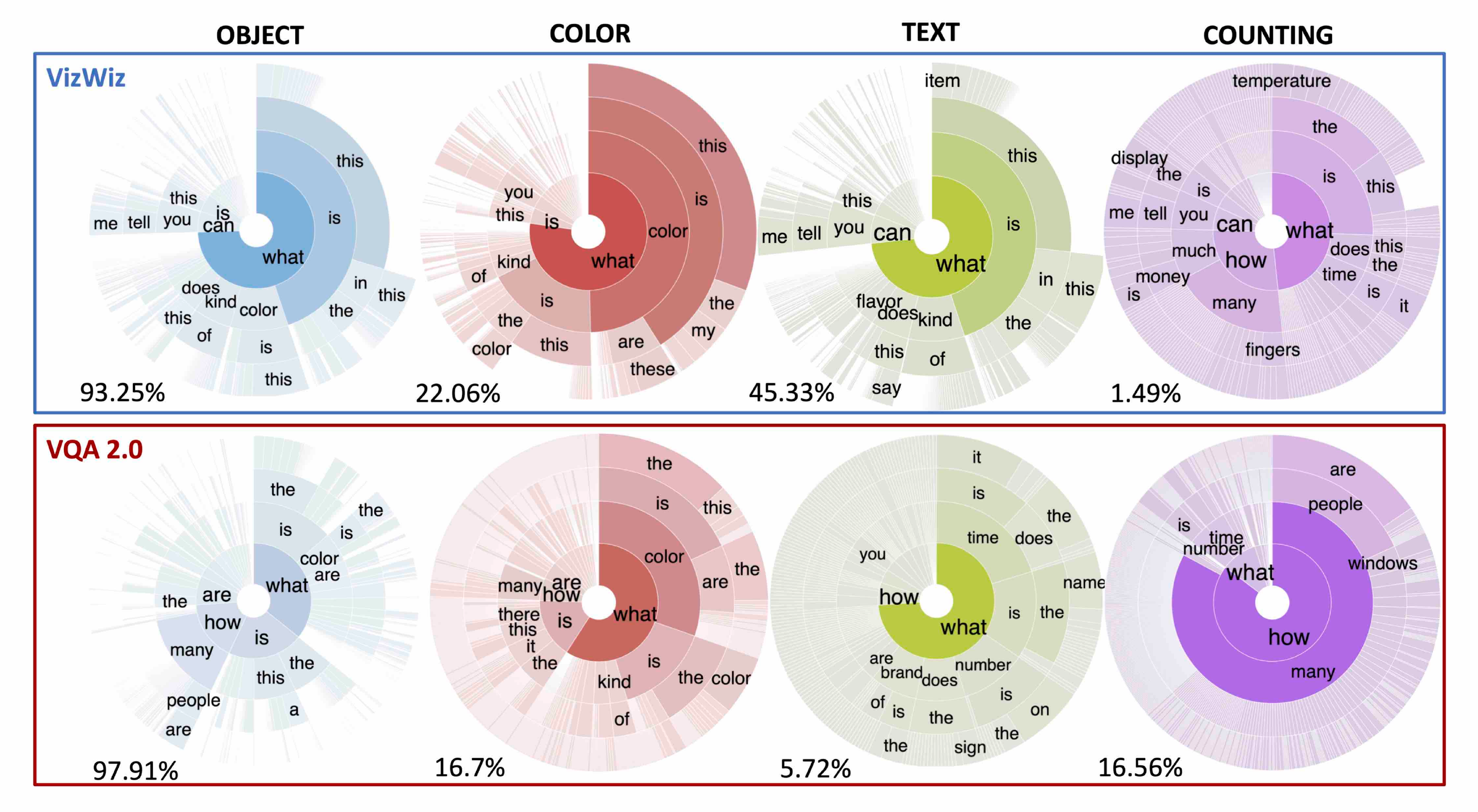}
    \caption{Shown is the percentage of visual questions needing each skill to answer the questions in VizWiz and VQA2.0, paired with a sunburst diagram showing the first four words of all questions for each skill-dataset combination.  For the sunburst diagrams, the order of rings represents the order of words in the question, with the innermost ring representing the first word, and the arc size proportional to the percentage of questions with that wording. The level of color saturation for a word is relative to the frequency of the word, with greater saturation indicating the word is more common.}
    \label{fig:sunburstplots}
\end{figure*}

We next examined the types of questions that are asked for each skill.  To do so, we analyzed the frequency that questions begin with different phrases.  Results are shown in Figure~\ref{fig:sunburstplots} as a sunburst plot for each of the eight skill-dataset combinations for the first four words of each question.  We also quantified the most frequent words in the questions for each skill category.  Table \ref{tab:most_freq_words} shows the results in descending order of frequency, excluding pronouns and expressions such as ``please'' and ``thank you.''

\begin{table}[t!]
\centering
\resizebox{\columnwidth}{!}{%
\begin{tabular}{|l|l|l|}
\hline
\textbf{Skill} & \textbf{Dataset} & \textbf{Most frequent words} \\ \hline
\multirow{2}{*}{\begin{tabular}[c]{@{}l@{}}Object\\ recognition\end{tabular}} & VizWiz & color(s) (12\%), tell (7\%), kind (7\%), picture (4\%), shirt (4\%) \\ \cline{2-3} 
 & VQA2.0 & man (23\%), many/much (15\%), color(s) (7\%), people (6\%), picture (5\%) \\ \hline
\multirow{2}{*}{\begin{tabular}[c]{@{}l@{}}Color\\ recognition\end{tabular}} & VizWiz & color(s) (39\%), shirt (15\%), kind (3\%), box (1\%), bottle (1\%) \\ \cline{2-3} 
 & VQA2.0 & color(s) (43\%), man (12\%), wearing (5\%), red (5\%), person (3\%), white (3\%) \\ \hline
\multirow{2}{*}{\begin{tabular}[c]{@{}l@{}}Text\\ recognition\end{tabular}} & VizWiz & box (5\%), flavor (4\%), bottle (4\%), product (4\%), item (3\%) \\ \cline{2-3} 
 & VQA2.0 & time (21\%), clock (13\%), number (16\%), sign (11\%), brand (8\%) \\ \hline
\multirow{2}{*}{Counting} & VizWiz & many/much (32\%), number (15\%), temperature (10\%), time (8\%), money (5\%) \\ \cline{2-3} 
 & VQA2.0 & people (19\%), many/much (18\%), picture (12\%), number (5\%), visible (4\%) \\ \hline
\end{tabular}%
}
\caption{The top 5 most frequent words in visual questions that belong to each of the four vision skills. Numbers in parentheses indicate the percentage of questions within each category that contain the word.}
\label{tab:most_freq_words}
\end{table}

We observe a considerable difference in the type of questions asked across the two datasets for object recognition.  Many VizWiz questions are open-ended and vague, while many VQA2.0 questions identify the object of interest in the image as part of the question (first column in Figure~\ref{fig:sunburstplots}).  For example, approximately $26\%$ of all VizWiz's visual questions are variants of the question ``What is this?'', including the questions ``What is it?'', ``What is that?'', or ``Can you tell me what this is, please?''  In contrast, the questions in VQA2.0 often contain extra language-based guidance describing the content of interest in the image, such as ``Are these trains moving?'', ``Is there celery?'', and ``A person with what kind of job would ride in this vehicle?''.  We attribute this difference partially to the different origins of the two datasets: VizWiz questions originate from people who could not see the images, while VQA2.0 questions come from sighted crowd workers who were shown the images as a prompt to help them contrive questions that might stump a smart robot.  

For text recognition and counting, we also observe different trends in the types of questions asked.  For text recognition, we again found that VizWiz questions seem to be more broad and ambiguous than those from VQA2.0. Many VizWiz questions are variants of ``What is this?'', while many VQA2.0 questions start with phrases such as ``what is the name...?'', ``what number...?'', and ``what time...?''\footnote{A similarity of both datasets is that a few questions appear to be mistakenly tagged as counting when they involve numbers in some way; e.g., ``What time is it?''.  This suggests that there is some noise in the skills labeling process.  These questions account for less than $1\%$ of the questions in both datasets.} (third column in Figure~\ref{fig:sunburstplots}).  We again attribute this difference to the way the questions were generated---VQA2.0 questions come from people who first observed the images, while people who are blind could not know what content was in the images that they asked questions about.  For counting, about 85\% of these questions in VQA2.0 start with ``how many...?''  (fourth column in Figure~\ref{fig:sunburstplots}).  Unsurprisingly, the keywords ``many'' and ``much'' are highly correlated with the counting skill (Pearson's $r(5,034) = 0.85$, $p < .001$).  In contrast, the phrases ``how many...?'' and ``how much...?'' constitute only about one quarter of the counting questions in VizWiz (fourth column in Figure~\ref{fig:sunburstplots}).  Moreover, we observe for counting questions that real users of VQA services (i.e., VizWiz users) commonly try to read about daily ``items/products'' contained in ``boxes'' and ``bottles'' while the AI community seems to be focusing on a narrower, less practical problem of reading ``numbers'' on ``clocks'' to learn the ``time'' or reading ``signs'' and ``brand'' names (Table \ref{tab:most_freq_words}).  These findings further highlight a mismatch between the interests of real users of VQA services and the AI community's focus in developing VQA algorithms.

In contrast, we observe similar trends across both datasets for the color recognition questions.  Often ``color'' is \textit{explicitly} mentioned in the first four words of the question (second column in Figure~\ref{fig:sunburstplots}).  Moreover, it is the most common word in color recognition questions, present in 39\% of VizWiz and 43\% of VQA2.0 (Table \ref{tab:most_freq_words}) and highly correlated with visual questions requiring color recognition (i.e., Pearson's $r(22,226)= .75$, $p < .001 $ for VizWiz; Pearson's $r(5,034) = .60$, $p < .001 $ for VQA2.0).  Many questions also \textit{implicitly} request color information.  For example, in Figure~\ref{fig:vq_examples}, it takes common sense about expected colors to answer questions such as ``Is it okay to cross the street?'' (VQA2.0) and ``Does that [surface] look clean?'' (VizWiz).  Across both datasets, many questions are also concerned with clothing.  For example, roughly 15\% of VizWiz color recognition questions ask about a ``shirt,'' and 5\% of VQA2.0 color recognition questions ask about ``wearing'' something (Table \ref{tab:most_freq_words}).  In addition, roughly $6\%$ of VizWiz's color recognition questions are one of the following: ``What color are these shirts/pants/shoes?''.  The difference between the datasets appears to lie in that VQA2.0 manifests more descriptive color recognition questions compared to VizWiz.  Specifically, two of the most frequent words in VQA2.0, ``red'' and ``white,'' are colors, whereas the top words from VizWiz questions demonstrate an interest in \textit{learning} about such attributes, including for a ``shirt,'' ``box,'' and ``bottle'' (Table \ref{tab:most_freq_words}).

\subsection{Discussion}
 Our analysis of the skills needed to answer visual questions reveals similarities and differences for the two studied use cases.  In summary, while real VQA service users are mostly concerned with learning about the text and color of common objects, the AI community more strongly emphasizes learning about the type, count, and color of objects in images showing 80 object categories.  We provide additional finer-grained analysis about the types of questions solicited for each skill type as well as other needed skill types in the Appendix.  
 
 Our findings serve as a valuable foundation for creating future large-scale datasets for machine learning that are more inclusive (i.e., addressing blind people’s interests) and general-purpose (i.e., work well for a greater diversity of scenarios).  It reveals the AI community's ``blind spots'' that resulted from excluding consideration of real users of VQA services when building today's mainstream dataset.  At present, we can expect automated VQA services will consistently fail for real users, especially for any text recognition tasks (which is nearly half of their requests!).  We offer our findings as a valuable foundation for demonstrating what types of visual questions need to be included/added in future datasets for the machine learning community.  Our work also reinforces the importance of participatory design~\cite{harrington2019DeconstructingCommunityBasedCollaborative} in creating such crowdsourcing tasks and motivates the importance of changing the status quo, such as by engaging real users with crowd workers to guide them to generate practically relevant data (e.g., extending the work of Venkatagiri et al.~\cite{venkatagiri2019GroundTruthAugmentingExpert}).

 Our findings are also valuable in guiding the design of future VQA algorithms.  Our skills labels reveal the internal reasoning that is vital for building successful algorithms.  It is important for them to include multiple stages, such as recognizing color and text towards solving counting.  The AI community's efforts to develop attention-based VQA algorithms \cite{lu2016hierarchical,fukui2016bilinear-pooling,yang2016stacked}---finding the object(s) of interest in an image as an intermediate step towards answering a visual question---is a promising step in this direction, since object recognition is nearly always necessary for answering visual questions.  Going forward, algorithms could be trained to exploit the other skill correlations in order to decide which skills are needed when in a multi-stage process. 
\section{Difficulty of Skills for Humans and Computers}
We next examine what makes it difficult to answer visual questions.  To do so, we analyze the ability of crowd workers and readily available, off-the-shelf algorithms (aka - computers) to address each skill.  This offers valuable insights regarding what skills humans need better training for and which algorithmic reasoning skills most need improvement.

\subsection{Difficulty for Humans}
Although there is a limited precedence for how to measure the difficulty of a visual question, we employ two recently posed measures~\cite{yang2018VisualQuestionAnswer,bhattacharya2019does} for this purpose. They are computed with respect to each skill for both datasets. 

\paragraph{Difficulty Measures}
We characterized difficulty with two measures based on analyzing answers provided by independent human respondents: the heterogeneity of the answers as well as the reasons why they are heterogeneous.  We describe both measures below.
\begin{itemize}
    \item \textbf{Answer Entropy}: Inspired by prior work~\cite{yang2018VisualQuestionAnswer}, we computed the entropy of the 10 crowdsourced answers for each visual question as follows: 
\begin{equation}
 E=\sum_{i=1}^{N}-p_{i} \log p_{i}
\end{equation}
\noindent
where $N$ represents the number of unique answers and ${p_i}$ represents the proportion the $i$-th answer occupies in the 10 answers. When the 10 people independently offer the same answer, the score is 0. This scenario reflects when the question asks for common sense information, and so we assume is easier to answer. At the other extreme is when the 10 people each offer a different answer and so the score is $3.32$. In this case, the correct answer is likely not common sense since a large diversity of answers arise.  To avoid measuring trivial syntactic differences, we followed the pre-processing steps proposed by prior work~\cite{vqaEvaluation} to make all letters lowercase, remove periods except those in decimals, remove articles (``a'', ``an'', ``the''), add apostrophes to contractions that lack them (e.g., ``cant'' - ``can't''), convert numbers that are spelled out to digits, and then use a space character to replace all punctuation except apostrophes and colons.
    
\item \textbf{Reasons for Answer Differences}: We employed the Answer-Difference dataset~\cite{bhattacharya2019does}, which indicates which among 10 reasons explain why answers are different for each visual question in both VizWiz and VQA2.0.  Six of the reasons stem from the nature of the visual question: low quality image (LQI), answer not present in the image (IVE), invalid question (INV), question needing special expertise (EXP), ambiguous question (AMB), and subjective question (SBJ).  The remaining three reasons arise because of the answers, including that they can be synonyms (SYN), describing the same things with different levels of granularity (GRN), or spam (SPM). There is also an ``other'' (OTH) category to catch any fall-through.  The Answer-Difference dataset~\cite{bhattacharya2019does} was created by showing crowd workers each visual question with 10 previously crowdsourced answers and asking them to identify all the reasons among the aforementioned 10 reasons for why there are observed answer differences.  We employ the resulting metadata, which includes 10 labels per visual question indicating which reasons for answer-difference are relevant, to support our analysis. We focus on examining the extent that these reasons for answer differences correlate with each skill.  
\end{itemize}

\begin{figure}[b!] 
 \centering
    \subfloat[VizWiz]{\includegraphics[width=0.3\columnwidth]{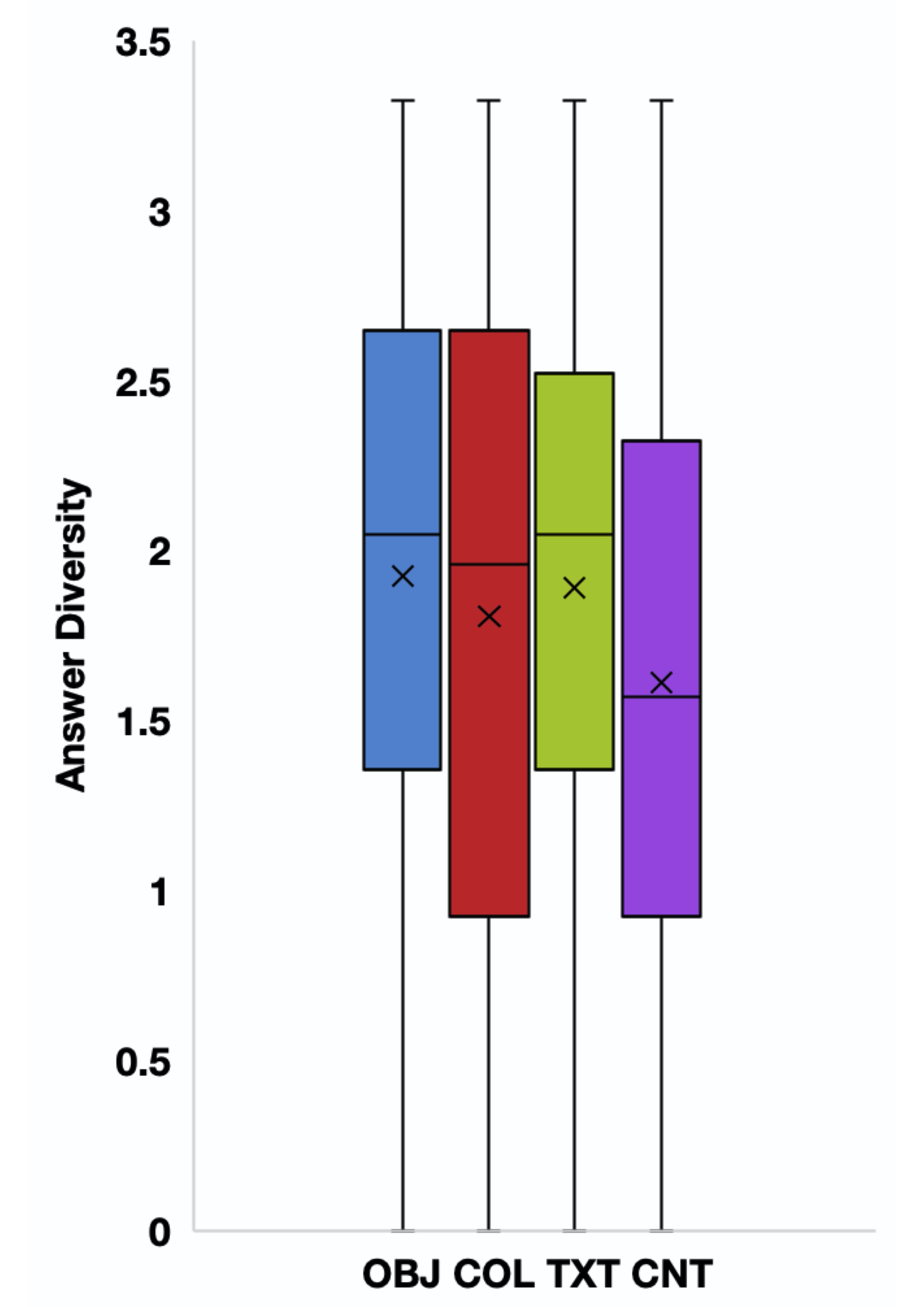}}
    \subfloat[VQA2.0]{\includegraphics[width=0.3\columnwidth]{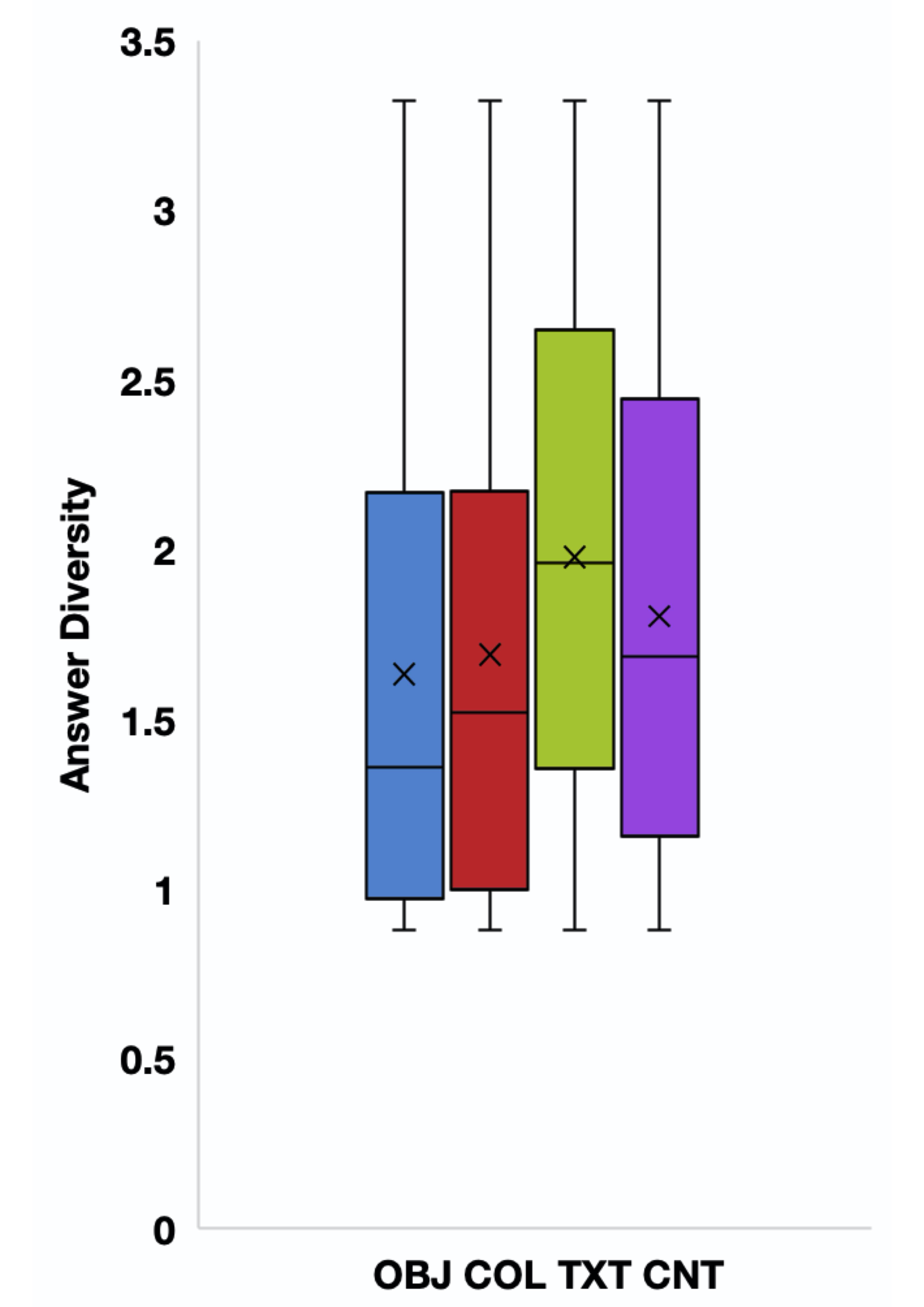}}
\caption{The difficulty of different skills for humans, computed using the answer entropy measure for two datasets.  For each box, the horizontal line denotes the median  and the cross mark denotes the mean. The edges denote the 25th and 75th percentiles scores, and the whiskers extend to the most extreme data points that are not considered outliers.} 
\label{fig:answerDiversity}  
\end{figure}

\begin{figure*}[t!] 
 \centering
    \subfloat[VizWiz]{\includegraphics[width=.5\columnwidth]{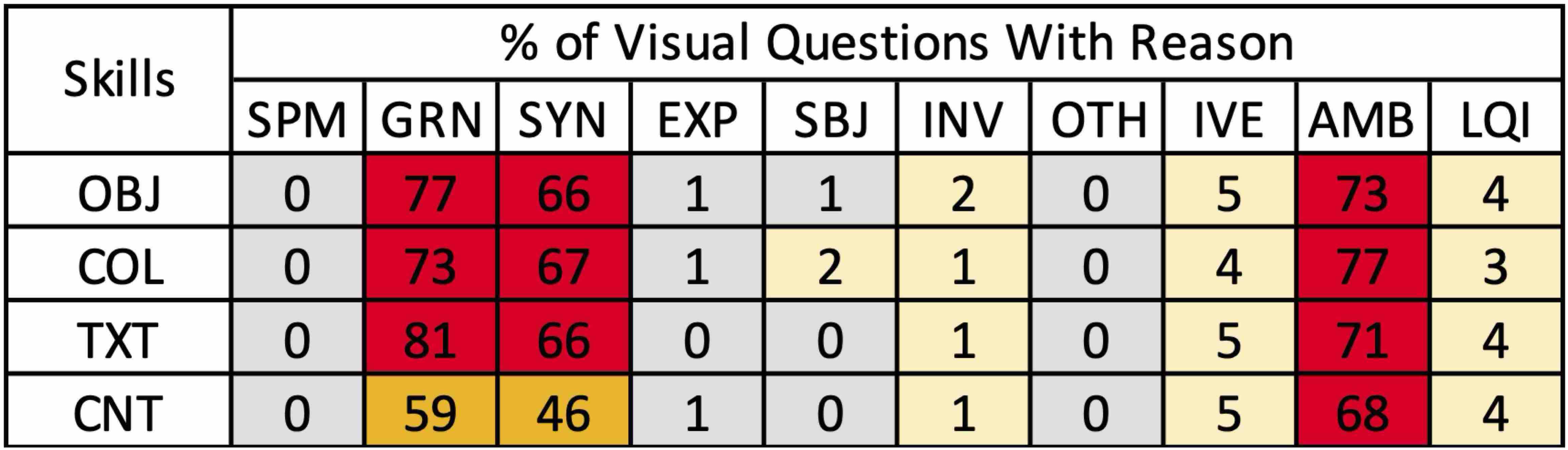}}
    \subfloat[VQA2.0]{\includegraphics[width=.5\columnwidth]{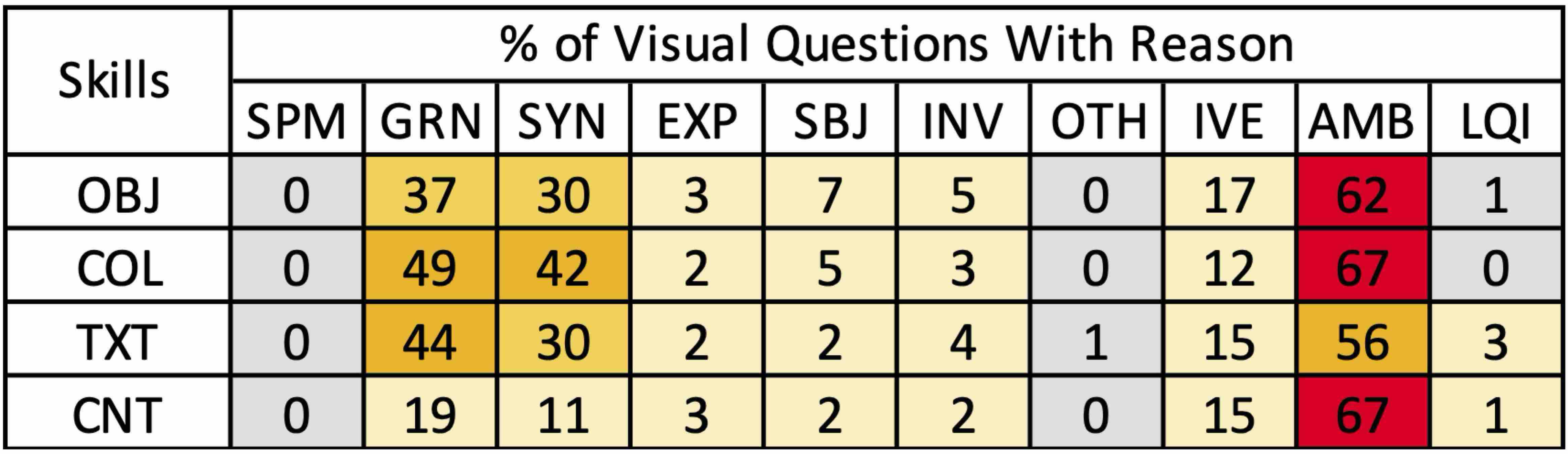}}
\caption{Percentage of visual questions that led to answer difference that are due to various reasons for each skill-dataset pair.  A red cell indicates that more than $60\%$ of the questions needing the skill are correlated with the corresponding reason. A {grey} cell indicates that less than $1\%$ of the questions needing the skill are correlated with the corresponding reason (best viewed in color).
} 
\label{fig:reasonDistribution}  
\end{figure*}

\paragraph{Results}
Results for the entropy scores and corresponding reasons for observed answer differences are shown for each skill-dataset pair in Figures \ref{fig:answerDiversity} and \ref{fig:reasonDistribution} respectively.
 
 For text recognition, we observe that the difficulty trend is similar across both datasets, with it typically being a more difficult skill than the other skills.\footnote{Statistical testing analysis that reveals the significance of differences are shown in the appendix.}  For example, mean entropy score for this skill in VQA2.0 is 1.98, which is 0.18 greater than the next highest score of 1.80 for the counting skill (Figure~\ref{fig:answerDiversity}).  The most common reason for this difficulty is that the visual questions are ambiguous (i.e., AMB for 71\% in VizWiz and 56\% in VQA2.0).  This ambiguity in turn leads to different interpretations about how to answer questions, including what level of detail to provide (i.e., GRN for 81\% in VizWiz and 44\% in VQA2.0) and which words/phrases to use (i.e., SYN for 66\% in VizWiz and 30\% in VQA2.0).  We hypothesize that the difficulty stems from text in the images often being long and descriptive.  We suspect that people struggled with ambiguity (AMB) in how to extract or summarize information in the text and so varied in the amount of detail they provide (GRN) and how they summarize their findings (SYN).  

There is considerable disparity in the difficulty of visual questions that require object recognition and color recognition skills across the two datasets: the difference in the mean entropy scores is 0.29 for object recognition and 0.12 for color recognition.\footnote{Statistical testing analysis that reveals the significance of differences are shown in the appendix.} For both skills, questions from VizWiz tend to be more difficult for humans. Compared to VQA2.0, VizWiz's answer differences more often occur because answers have varying levels of detail (GRN) or synonyms (SYN).  These findings highlight that the difficulty lies partly in knowing how to identify the most suitable answer when many options are possible.  Other possible reasons for this increased difficulty are that (1) VizWiz's object recognition and color recognition questions more often require other skills to answer them (as shown in Figure~\ref{fig:skill_comb_dist}), and (2) VizWiz questions are less precise than those from VQA2.0, where people asking questions saw the image as a priori.  

Altogether, these findings offer valuable insights into what skills training may be most impactful to achieve crowd consensus when collecting answers from remote, crowdsourced humans.  We anticipate that focusing on training for skills that are consistently more challenging (i.e., text recognition) would be most beneficial.  For text recognition, we believe that greater guidance in how to summarize text when answering questions would make this skill type less difficult for humans.  In contrast, for skills that span a large range of difficulty levels for humans (i.e., object recognition and color recognition), we believe extra work is desirable to detangle what conditions makes each skill easy versus difficult.  Such knowledge, in turn, would serve as a valuable foundation to reveal what training for subsets of these skills may be most effective to achieve crowd consensus.

\subsection{Difficulty for Computers}
We also evaluated the difficulty for computers to predict an answer correctly with respect to the different skills.

\paragraph{Methods}
We benchmarked off-the-shelf methods that are designed to solve the specific skills discussed in this study.  We evaluated a total of eight methods spanning the four skill categories. For each skill category, we focused only on algorithms that are related to the skill. The methods are described below.

For object recognition, we evaluated three methods: 
\begin{itemize}
    \item \textbf{Objects}: objects recognized by the YOLO object detector (version 3)~\cite{yolov3}.
    \item \textbf{Descriptions}: a one sentence long description of an image generated by the Microsoft Azure Vision API.
    \item \textbf{Tags}: a list of objects, living things, scenery, and actions found in the image by the Microsoft Azure Vision API.
\end{itemize}

For color recognition, we evaluated Microsoft Azure's ``Descriptions'' and ``Tags'' as well as the following two methods:
\begin{itemize}
    \item \textbf{Dominant Colors}: the dominant colors extracted using the Microsoft Azure API.
    \item \textbf{Dominant Colors (Seg)}: the dominant colors recognized by identifying the most common color (using the xkcd palette~\cite{xkcdPa}) for each segmented region detected by the YOLO object detector (version 3)~\cite{yolov3}.
\end{itemize}

For text recognition, we evaluated Microsoft Azure's ``Descriptions'' and ``Tags'' as well as the following two methods: 
\begin{itemize}
\item \textbf{OCR Text}: text recognized with Microsoft Azure OCR API.
\item \textbf{Handwritten Text}: text recognized with Microsoft Azure Recognize Text API.
\end{itemize}

Likewise for counting, we evaluated Microsoft Azure's ``Descriptions'' and ``Tags'' as well as the following method: 
\begin{itemize}
\item \textbf{Num}: count of objects detected by YOLO (version 3)~\cite{yolov3}.
\end{itemize}

We offer this list as an initial set of baselines to begin to understand the difficulty of each skill for widely-available, off-the-shelf methods that pertain to each skill.

\paragraph{Evaluation Metric}
To measure the difficulty for computers to answer visual questions, we employed the standard VQA evaluation metric used by the AI community~\cite{vqaEvaluation}.  It uses human answers as the ground truth while acknowledging that different humans may offer multiple valid answers. It gives partial credit if the computer comes up with an answer that matches at least one of the 10 human-generated answers, and full credit to any computer-generated answer that at least 3 of 10 people have offered. To avoid penalizing algorithms for minor, non-semantic differences between the predicted answer and human-benchmarked answers, we deem an answer as a match if the algorithm's prediction contains the answer (rather than employing exact string match).  Formally, this metric is defined as:
\vspace{-0.25em}
\[Acc(ans) = \min \{ \frac{{\# {\rm{\ humans'\ answers\ contained\ in\ ans}}}}{3},1\} \]

\paragraph{Baseline - Upper Bound}
We also report what is the best human performance possible. To do so, for each visual question, we identified the most popular answer from the 10 available answers.  We call this the ``Human Majority Answer.''  Measuring the performance of the ``Human Majority Answer'' for all visual questions with the evaluation metric above yields the upper bound of performance that is possible for a dataset.

\begin{figure}
    \centering
    \includegraphics[width=\textwidth]{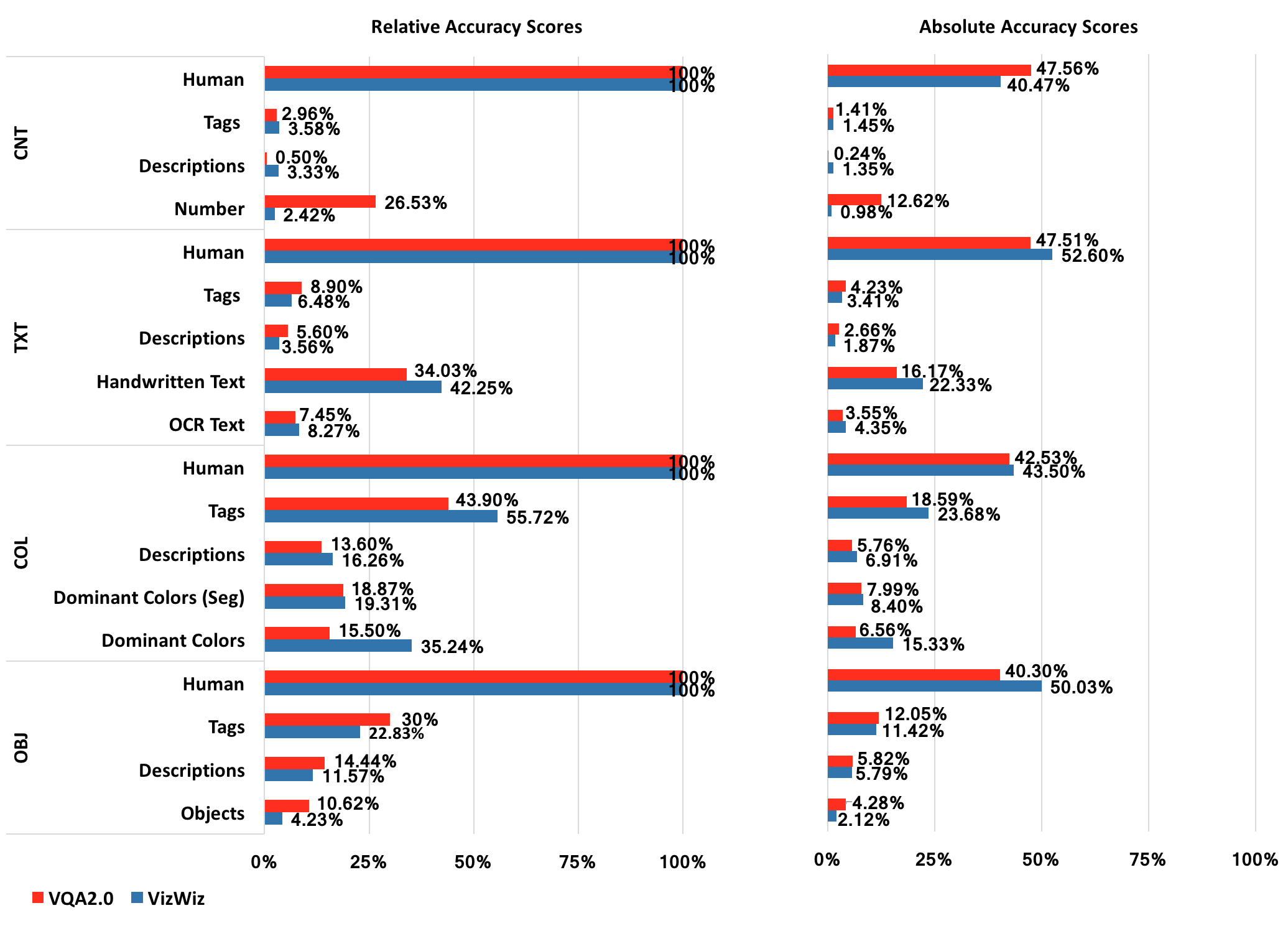}
    \caption{Relative (left) and absolute (right) accuracy scores of computer-generated answers to visual questions compared to the best possible performance from humans (100\% relative accuracy). Results are shown for two datasets.  Algorithms recognizing text and color achieved relatively high accuracy scores for questions that require such skills. In contrast, algorithms performed poorly for questions involving object recognition.}
    \label{fig:vqa_accuracy}
\end{figure}

\paragraph{Results}
Results are shown in Figure \ref{fig:vqa_accuracy}.   

We observe that algorithms perform closest to human performance for two skills: color recognition and text recognition. For color recognition, the top-performing method (i.e., ``Tags'') resulted in an accuracy that is approximately $56\%$ and $44\%$ of human performance for VizWiz and VQA2.0 respectively.  For text recognition, the top-performing method (i.e., ``Handwritten Text'') achieved an accuracy that is approximately $42\%$ and $34\%$ of human performance for VizWiz and VQA2.0 respectively.  This finding is especially exciting since so many of real users' visual questions require these two skills of text recognition and color recognition.

A surprising aspect of the results is that the algorithms perform better for the visual questions from real users (i.e., VizWiz) than for those in the fabricated VQA dataset (i.e., VQA2.0), with 12 and 8 percentage point differences for color recognition and text recognition, respectively.  We found this surprising since algorithms are (arguably) rarely trained on data originating from people who are blind, given the scarcity of such data.  One possible reason for this is that real users tended to submit images where they zoomed into the content that was the focus of the question they asked about, thereby making it easier for algorithms to recognize the text and color.  Additionally, we found that a higher proportion of visual questions in VQA2.0 dataset have images that either do not contain an answer to the question (IVE, INV) or are subjective (SBJ), thereby also making them harder to answer.  

In contrast, we found that methods that directly recognize objects in an image often fail to provide accurate answers to the visual questions.  The top-performing method (i.e., Tags) achieves approximately 23\% and 30\% of human performance in VizWiz and VQA2.0 respectively.  We hypothesize this poor performance is largely because many questions require more skills than simply recognizing the object, as discussed in the previous section.  We suspect the better performance on VQA2.0 is in part because many more of those visual questions require only object recognition.

Visual questions that need the counting skill also often fail to correctly answer visual questions.  Performance is better for VQA2.0, for which we observe an accuracy of approximately $27\%$ of human performance for the top-performing method.  We again attribute the overall poor performance to the need to perform multiple skills in order to count, as discussed in the previous section.  

We conclude that the skill difficulty levels for computers, from easy to hard, are color recognition, text recognition, object recognition, and counting.  This is true when observing the performance of the top-performing methods both relative to human performance as well as in absolute terms (Figure~\ref{fig:vqa_accuracy}).  This finding is consistent across both datasets, and so generalizes across a large diversity of data.  These findings underscore the limits of automated visual assistance services today, which is important for setting users' expectations from such technologies.

\subsection{Discussion}
Our findings reinforce the importance of improving the clarity of the VQA task for humans.  With it being well-known that different people can produce a heterogeneity of answers for the same visual question~\cite{antol2015VqaVisualquestion,gurari2017CrowdVergePredictingIf,yang2018VisualQuestionAnswer}, we offer our findings as a valuable step in guiding what skills training could be most impactful to achieve crowd consensus when collecting answers from remote, crowdsourced humans.  In particular, at present, humans' answer variance is greatest for text recognition, followed by object recognition, color recognition, and counting.  This ordering highlights the overall prioritization of training across the skills.  We believe our findings could also be of broader benefit to any service provider that train humans to answer visual questions from people who are blind, such as Aira~\cite{HomeAiraAira}, BeMyEyes~\cite{bemyeyes}, and BeSpecular~\cite{BeSpecular}.

Our findings also offers insights into what are ``easy'' versus ``hard'' skills for widely-available automated methods.  Our findings reveal that computers perform considerably better for some skills than others.  Still, we observe that there is considerable room for improvement in computers to mimic human performance, with none of the skills manifestly trivial.  
\section {Predicting What Skills Are Needed}
We next examined a novel problem of predicting directly from a visual question which skills are needed to provide an answer.  We posed the task as a multi-label classification problem where, given an image-question pair, a predictor decides which of the skills are needed to answer that question.  We excluded object recognition from our prediction task since it is nearly always needed and so is a trivial prediction task.

A skill predictor could be beneficial in practice for several scenarios. For example, it can be used in a ``swiss-army-knife'' VQA framework, in which identified relevant skills will help in dynamically determining how to route visual questions to the appropriate agent, human or computer, based on contingencies such as cost or difficulty. 

\subsection{Method}
We use the top-down model \cite{anderson2018bottom} for skill prediction, given its strong performance in solving VQA tasks. It takes as input image features encoded by ResNet-152 \cite{he2016deep} and a paired question encoded by a GRU cell. A top-down attention module computes a weighted image feature from the encoded question representation and the input image features. The weighted image feature is coupled with the question representation by element-wise multiplication. Eventually, this coupled feature is processed by two fully connected layers to predict the skill.  The final layer is a sigmoid function that predicts the probability for each skill.  Consequently, this architecture behaves as a ``binary classifier'' because a sigmoid function solves the binary classification task for each skill within a single algorithm (i.e., performs multi-label rather than multi-class classification).

We trained the model with the Adam optimizer, using a batch size of 256 image-question pairs for 20 epochs for each dataset. We evaluated the model from each epoch on the validation set and selected the one with highest F1 score for use on the test set. 

\subsection{Evaluation}

\paragraph{Dataset}
We divided both skills datasets into approximately 65\%/10\%/25\% training, validation, and test splits.  All results below are reported for the test dataset.

\paragraph{Evaluation Metrics}
Since the text, color, and counting skills to be predicted have significantly imbalanced distributions in both datasets (Figure~\ref{fig:sunburstplots}), we use F1 score, which is the harmonic average of precision and recall scores, as the main metric to evaluate model performance. We also visualize the precision-recall curve for each trained model to demonstrate the trade-off between precision and recall scores.  Finally, we report the average precision, which is the area under the precision-recall curve, and recall scores.

\paragraph{Baselines}
We employ a set of naive classifiers that randomly guess which skills are needed.  This baseline represents the best performance that can be achieved with a trivial prediction scheme. 

\begin{figure}[th!]
    \subfloat[Trained on images and questions]{\includegraphics[width=0.5\columnwidth]{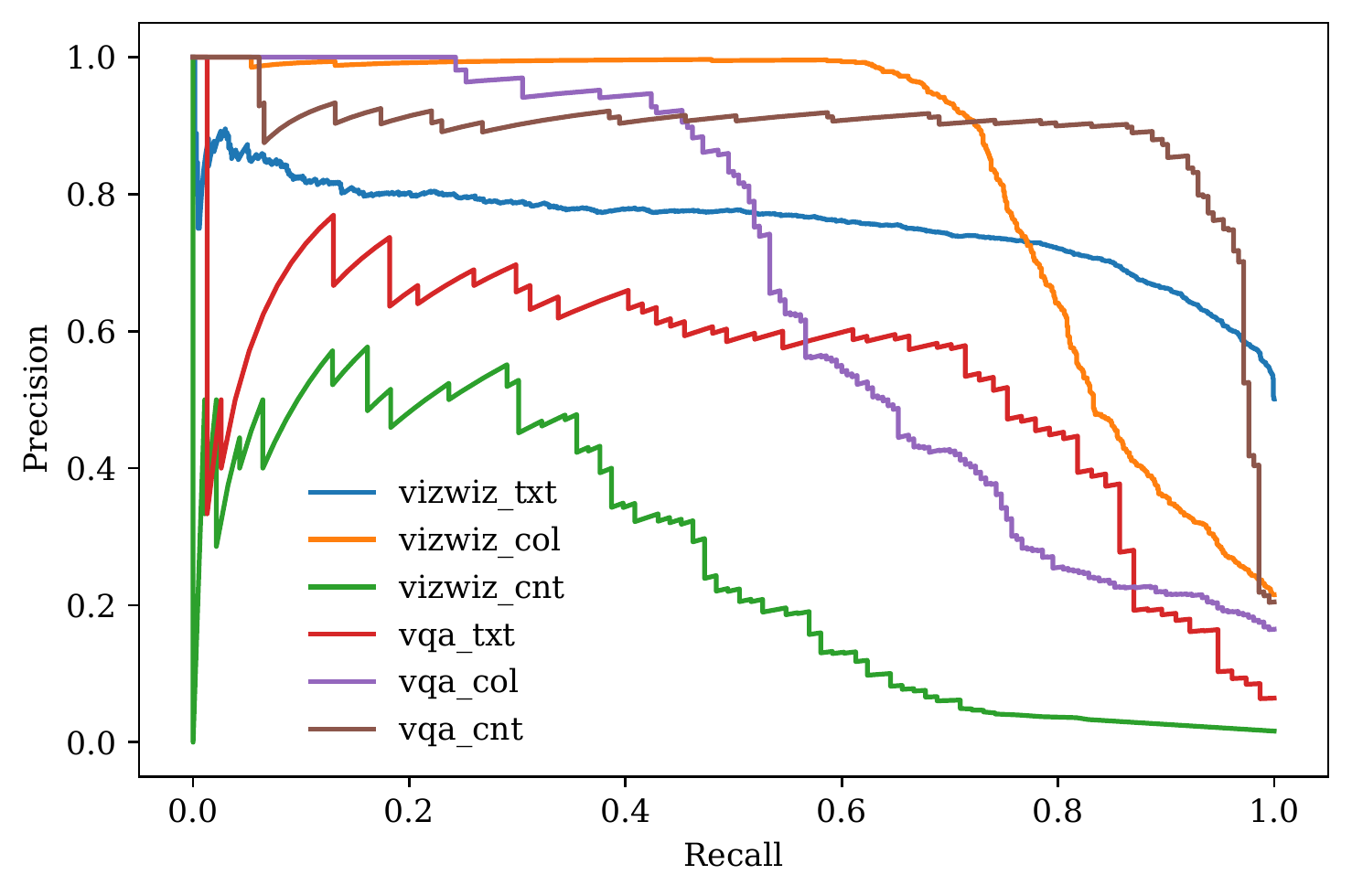}}
    \subfloat[Trained on questions only]{\includegraphics[width=0.5\columnwidth]{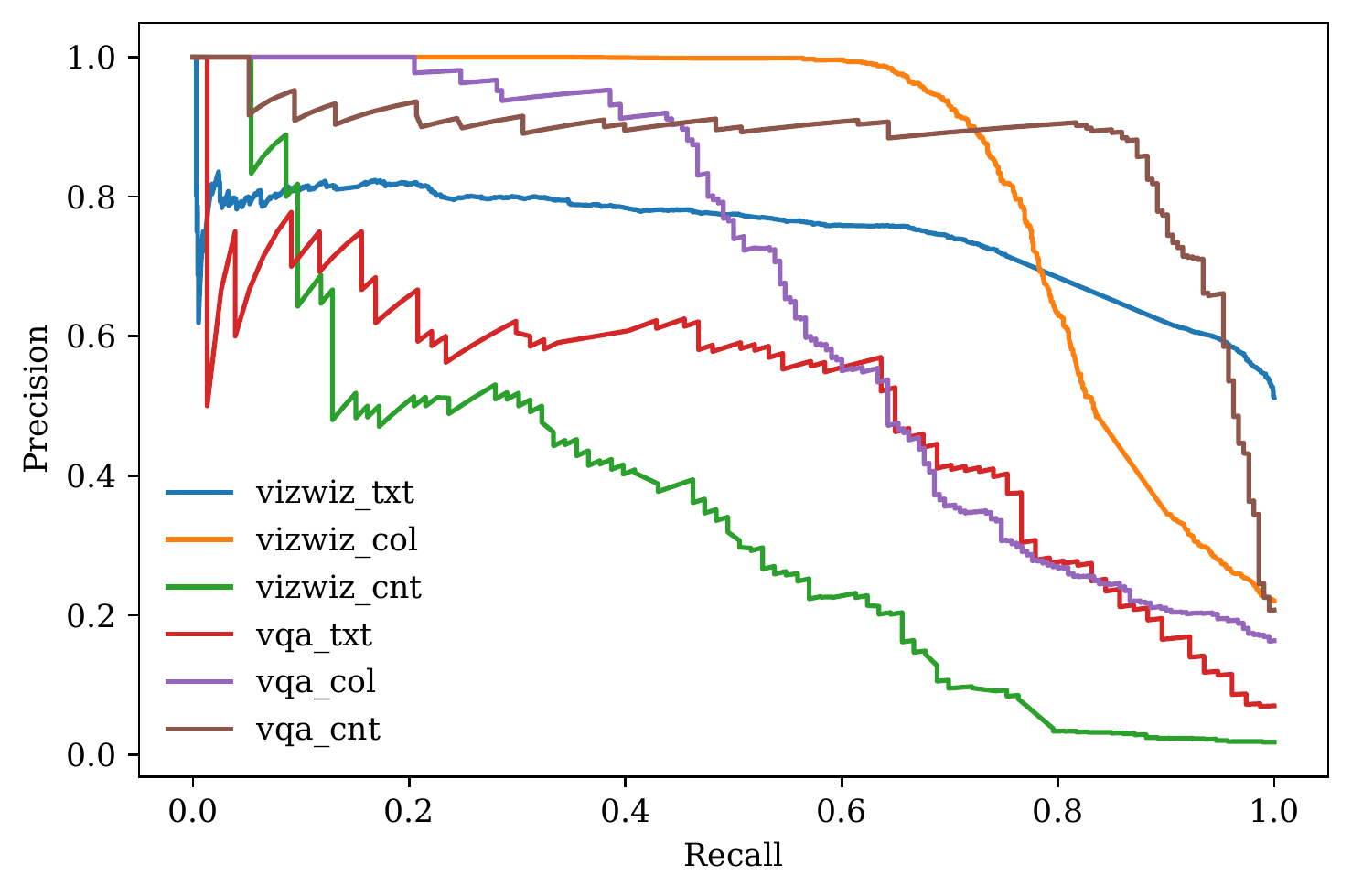}}
    \newline
    \subfloat[Trained on images only]{\includegraphics[width=0.5\columnwidth]{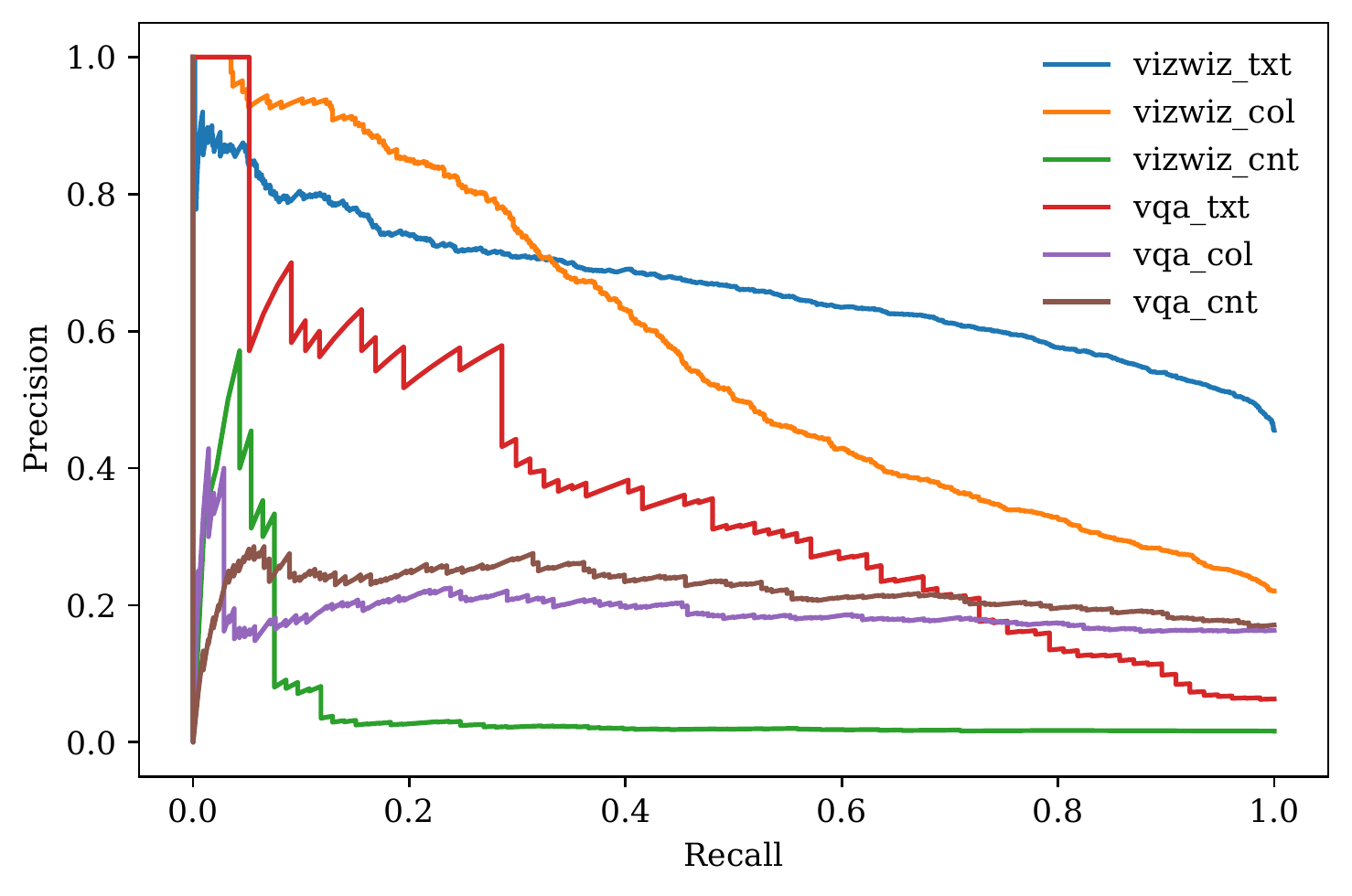}}
    \caption{Precision-recall curves for skill prediction using the top-down model.}
    \label{fig:precision_recall}
\end{figure}

\paragraph{Results}
Quantitative results indicating the performance of trained models are reported in Table~\ref{tab:skill_benchmarking} and visualized in Figure~\ref{fig:precision_recall}.  As observed, our models trained on both images and questions perform much better than what is possible today (i.e., random guessing) for all categories, with respect to all metrics (Table~\ref{tab:skill_benchmarking}). For instance, there is a boost in the F1 score for the color recognition skill of over 77\% for VizWiz (i.e., $21.8\%$ versus $79.3\%$) and $46\%$ for VQA2.0 (i.e., $16.6\%$ versus $63\%$).  For text recognition, we also observe a boost in the F1 score; specifically, 31.2\% for VizWiz (i.e., $45.2\%$ versus $76.4\%$) and $48.7\%$ for VQA2.0 (i.e., $16.6\%$ versus $63\%$).  The same trend is observed for counting, with a boost in the F1 score of 37\% for VizWiz (i.e., $1.6\%$ versus $38.6\%$) and $71.2\%$ for VQA2.0 (i.e., $16.5\%$ versus $87.7\%$).  Our results reveal it is possible to predict what vision skills are needed for a given image and associated question.  This is exciting to observe given that there is such a large diversity of questions and images in both datasets.

\begin{table}[b!]
    \centering
    \begin{tabular}{*8c}
    \toprule
         \multirow{2}{*}{Dataset} & \multirow{2}{*}{Skill} & \multicolumn{3}{c}{Random guessing} & \multicolumn{3}{c}{Top-down model} \\
    \cmidrule(lr){3-5}
    \cmidrule(lr){6-8}
        {} & {} & Precision & Recall & F1 & Avg. Precision & Recall & F1\\
    \midrule
        \multirow{3}{*}{VizWiz} & Text & 45.1 & 45.3 & 45.2 & 75.7/73.9/66.6 & 86.8/72.4/83.2 & 76.4/72.8/67.5 \\
        {} & Color & 21.5 & 22.1 & 21.8 & 85.0/84.8/56.8 & 67.6/69.3/40.1 & 79.3/79.8/49.0 \\
        {} & Counting & 1.6 & 1.5 & 1.6 & 26.2/33.9/5.1 & 35.5/28.0/8.6 & 38.6/36.6/8.7 \\
    \hdashline
        \multirow{3}{*}{VQA2.0} & Text & 6.0 & 5.7 & 5.8 & 54.6/42.6/36.2 & 50.6/57.1/28.6 & 54.5/46.3/36.0 \\
        {} & Color & 16.4 & 16.8 & 16.6 & 67.4/66.1/19.1 & 51.4/58.6/0.0 & 63.0/58.7/0.0 \\
        {} & Counting & 16.5 & 16.6 & 16.5 & 89.0/87.3/22.3 & 90.1/87.8/0.0 & 87.7/86.8/0.0 \\
    \bottomrule
    \end{tabular}
    \caption{Performance of skill prediction using random guessing and top-down model evaluated on VizWiz and VQA2.0 test datasets.  For the top-down model, three values from training on both images and questions, only questions,} and only images are reported for each metric, and the values of metrics are computed with the threshold 0.5.
    \label{tab:skill_benchmarking}
\end{table}

To gain greater insight into what makes the model predictive, we also report results from an ablation study where we trained the models on only the questions and only the images respectively. Results are reported in Table~\ref{tab:skill_benchmarking} and visualized in Figure~\ref{fig:precision_recall}. 

For the color recognition and counting skills, we observe that questions offer the greatest predictive power with question-only models, performing as well as those trained on both questions and images.  In contrast, image-based models perform quite poorly, as demonstrated by low values.  This is explainable in that models can learn to detect keywords, such as ``what color'' or ``is this red'' for the color skill and ``how many'' for the counting skill. However, models may not be able to learn much when only images are provided, since there may not be considerable difference in images' patterns when people aim to learn about an object's color or counts. 

For text recognition, we observe that both the image and the question are independently predictive.  We suspect the image-only model can learn to detect if text is present in images; i.e., it achieves $67.5\%$ in the F1 score for the text recognition skill in the VizWiz dataset, which is slightly lower than $76.4\%$ when trained on both images and questions.  For the text skill in the VQA dataset, we attribute the bad performance from training on images to the low percentage $5.72\%$ of that skill present in the dataset.  We suspect the question-only model can learn the common question structure when people seek assistance with reading text; i.e., it achieves $72.8\%$ in F1 score.  We observe that the models achieve the highest F1 scores for both datasets when trained on both images and questions, and so both inputs provide complementary meaningful information.

\begin{figure}[t!]
    \centering
    \includegraphics[width=\textwidth]{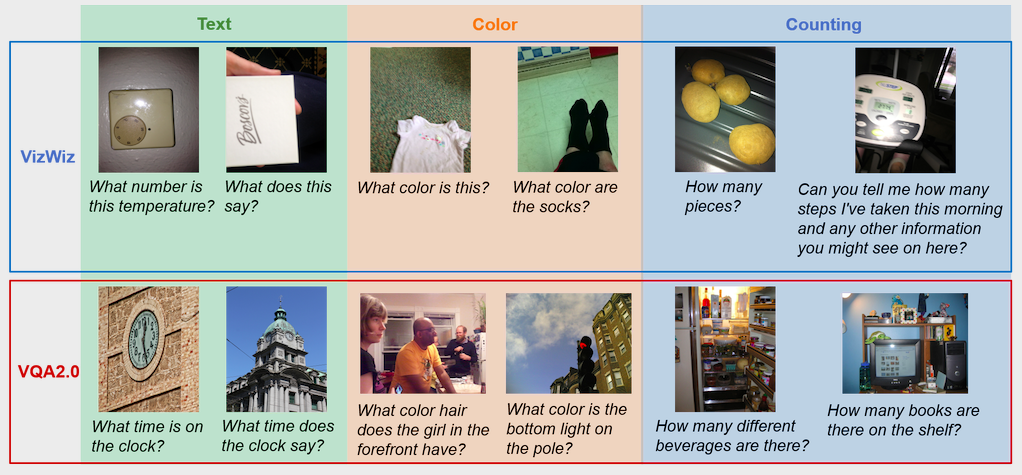}
    \caption{Top two most confident examples predicted by the top-down model for each skill and each dataset.}
    \label{fig:most_confident_predictions}
\end{figure}

For each skill and each dataset, we show the top two most confident examples predicted by the top-down models in Figure~\ref{fig:most_confident_predictions}.  We observe consistencies across both datasets with respect to the questions.  For instance, all four color recognition examples begin with the phrase ``what color.''  All four counting examples include the phrase ``how many.''  We conjecture that the prediction models rely on such key words/phrases in the questions to make predictions.  It's less evident to what extent the image content is relevant for making predictions, observing the considerable diversity of visual characteristics of the examples.  Specifically, the images range from showing complex scenes to a single object while also spanning indoor and outdoor settings.  We suspect these models behave like those reported by other researchers in the related VQA domain~\cite{antol2014zero, antol2015VqaVisualquestion, jabri2016revisiting, kafle2016answer} in that the questions are so important that often the relevant information can be predicted by looking at the question alone.  

\subsection{Discussion}
We demonstrate that an algorithm can achieve promising results in anticipating which skills are needed to answer visual questions. By automatically decomposing a complex task, the VQA task may join the ranks of other problems that are currently solved by divide-and-conquer efforts to offer a more efficient and accurate solution.  For example, content moderators on social media websites already employ automated methods to divide labor when deciding which content issues to address~\cite{jhaver2019Humanmachinecollaborationcontent}.

Accordingly, we believe our findings argue for the promise of numerous new research directions.  One possibility is a VQA service that dynamically routes each visual question to the best-suited agent, whether an algorithm, human, or combination of their efforts, based on the level of difficulty for each respective party. This design could integrate with existing applications such as Seeing AI \cite{seeingai} and EnvisionAI \cite{envisionai}, which currently enable users to choose visual assistance tools based on the skills they are seeking.  Future work will need to set the foundation to understand how the overhead of a manual solution (e.g., Be My Eyes \cite{bemyeyes}, VizWiz~\cite{bigham2010VizWiznearlyrealtime}) compares to that of a fully autonomous visual question answering system when designing such systems.  Cost-accuracy trade-offs would be useful in making such an assessment.  
\section{Limitations and Future Work}
Below we discuss several limitations of our work.  These have implications for potential fruitful areas for future work.

First, while the skill categories proposed in this paper were built off of the known skills discussed in existing literature and pilot studies, valuable future work is to refine and extend these skill categories into a collectively exhaustive and mutually exclusive taxonomy.  For example, text recognition could be further divided to sub-skills such as OCR (``What does this read?''), understanding (``For how long do I cook this in the microwave?'') and inference (``What does this box contain? Can you guess it by looking at it?'').  Another consideration will be to go beyond vision-based skills and consider the human-based reasoning skills that may also be required in order to answer visual questions (e.g., answering medical questions such as ``Is this infected?'').  Like prior work in another domain~\cite{chilton2013CascadeCrowdsourcingtaxonomy}, future work could look to the crowd to help in identifying the taxonomy of skills.  

Updates to the skill taxonomy could lead to numerous downstream benefits.  It could enable the design of more detailed guidelines for crowd workers to identify the most suitable skills, thereby improving the data quality.  It also could simplify the problem(s) that need to be solved by each algorithm if employing a swiss-army-knife VQA approach.

Another challenge remains of how to identify the most suitable metrics for assessing difficulty of visual questions for humans and computers.  While we pose metrics, we offer our work as a valuable starting point in starting to think about how to best achieve this.  

Finally, the skills prediction framework could be improved.  As is, our work poses this novel problem and offers a method that demonstrates the feasibility of automating this task.  We will publicly-share our new labelled dataset to stimulate future research.  A valuable area for extending this dataset will be in creating a more balanced dataset to represent all skills (e.g., the counting skill occupies only $1.49\%$ of the VizWiz dataset).  This could be help in supporting improved algorithm performance.
\section {Conclusion}
Our analysis reveals the unique information needs and challenges faced by real blind users of VQA services as well as the AI community in developing VQA algorithms, while also revealing the (mis)matches of the interests for these two use cases.  Our findings highlight the difficulty of generating answers for visual questions that belong to the different skills.  Finally, our findings demonstrate the potential of designing algorithms to automatically identify directly from a given visual question which vision skills are needed to return an answer.  The datasets and code are publicly shared at \texttt{https://vizwiz.org} to facilitate extensions of this work.

\section*{Acknowledgments}
We thank the crowd workers for providing the labels for our new dataset and the anonymous reviewers for their feedback.  This work is supported by National Science Foundation funding (IIS-1755593) and gifts from Microsoft. \\

\balance{}

\bibliographystyle{ACM-Reference-Format}
\bibliography{references,myReferences}

\section*{Appendix}
This document augments the content in Sections 3 and 4 of the main paper with the following:

\begin{itemize}
\item Dataset sampling approaches (supplements \textbf{Section 3.1})
\item Analysis of the consistency of skill labels collected from different crowd workers for each visual question (supplements \textbf{Section 3.1})
\item List of additional relevant skills that were suggested by crowd workers (supplements \textbf{Section 3.2})
\item Fine-grained analysis of the most common questions asked for each skill (supplements \textbf{Section 3.4})
\item Statistical testing for the comparisons of difficulty skill for humans (supplements \textbf{Section 4.1})
\end{itemize}

\subsection*{Dataset Sampling}
Expanding upon our discussion in \textbf{Section 3.1}, we elaborate on how we sampled the data from the VizWiz and VQA2.0 datasets.  We include all relevant VizWiz visual questions since our ultimate aim is to understand real users' needs and provide guidance to AI developers to align their efforts with real users' needs.  We chose a sample from the VQA2.0 dataset to support this analysis.  To do so, we randomly sampled approximately 5000 of the visual questions asked about 82,783 images, with 50\% stratified by question-type, and 50\% by answer-type.  These ``questions-types'' and ``answer-types'' were provided with the VQA2.0 dataset.  ``Question-type" refers to the kind of question that is asked based on up to the first 6 words of the question (e.g., ``how many..'', ``what is..'' etc.).  ``Answer-type'' refers to the kind of answers the questions are looking for: ``yes/no'', ``number'', or ``other''.  A question's type is defined to be the most popular option from the 10 labels assigned to the 10 answers per visual question.

\subsection*{Consistency of Skill Labels Collected from Crowd Workers}
\begin{figure*}[h!]
\centering
\includegraphics[width=0.9\textwidth]{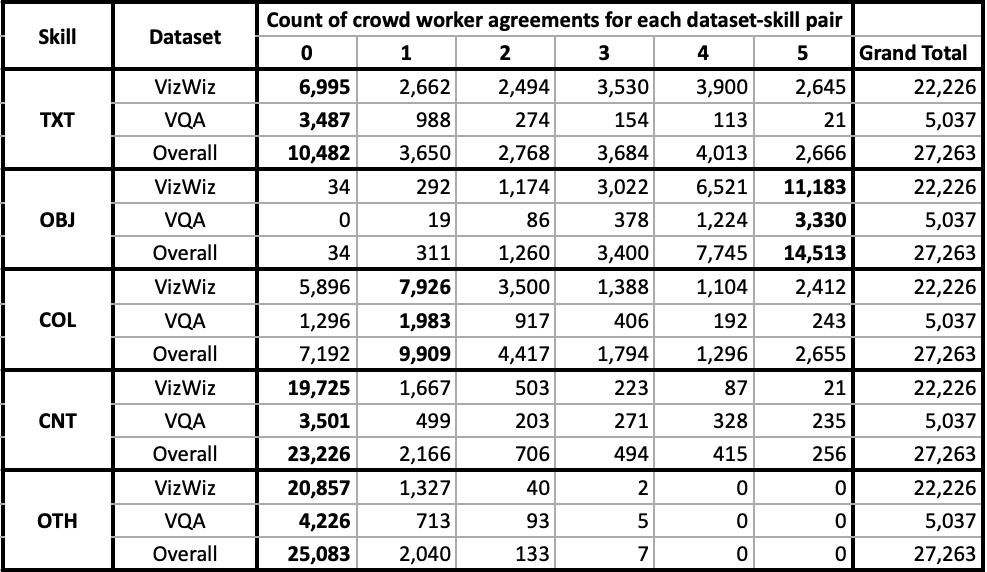}
\caption{Shown is the count of the number of votes from five crowd workers indicating the presence/absence of each skill for all images in both datasets.  For three of the four skills, there is most commonly unanimous agreement (i.e., count is 0 or 5) about the presence/absence of the skills.}
\label{fig_crowdAgreement}
\end{figure*}

\begin{figure*}[h!]
\centering
\includegraphics[width=0.9\textwidth]{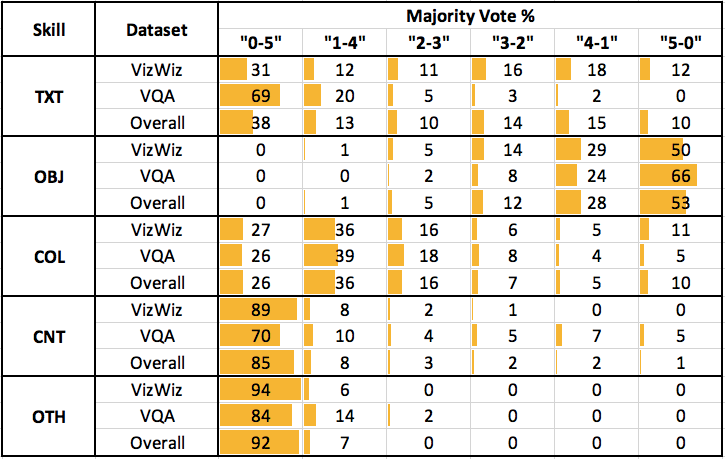}
\caption{Shown is the count of the number of votes from five crowd workers indicating the presence/absence of each skill for all images in both datasets.  For three of the four skills, there is most commonly unanimous agreement (i.e., count is 0 or 5) about the presence/absence of the skills.}
\label{fig_voteDistribution}
\end{figure*}

We tally how often 0, 1, 2, 3, 4, or all 5 crowd workers agree that each skill is needed for all images in both VizWiz and VQA2.0.  Results are shown in Figure~\ref{fig_crowdAgreement}.  Across both datasets for three of the four skills, most commonly there is unanimous agreement (i.e., count is 0 or 5) about the presence/absence of each skill.  We conduct complementary analysis to reveal what percentage of the labels arose from different voting distributions for both datasets and overall: 3-2, 4-1, versus 5-0.  Results are shown in Figure~\ref{fig_voteDistribution}.  

\subsection*{Other Relevant Skills}
We list here most of the top suggested skills submitted by crowd workers who chose the ``Other'' category, with the number of times they were suggested in parentheses: yes/no (86), reasoning and problem solving (67), situational awareness (53), unanswerable (38), spatial awareness/recognition (31), physical content recognition (25), food recognition (20), directional recognition (15), action/activity recognition (15), weather recognition (15), pattern recognition (14), opinion (13), judgement (12), location recognition (12), logo recognition (12), position recognition (11), direction (9), shape recognition (8), emotion recognition (7), facial recognition (7), inference (6), object orientation (6), brand recognition (5), place recognition (5), cooking skills (4), sports knowledge (4), symbol recognition (4), animal recognition (3), math skills (3), style recognition (3), time of day recognition (3), clean recognition (2).

\subsection*{Fine-grained Analysis of Most Common Questions for Each Skill}
\begin{figure}[htbp]
	\centering
	\subfloat[VizWiz]{\includegraphics[width=0.4\textwidth]{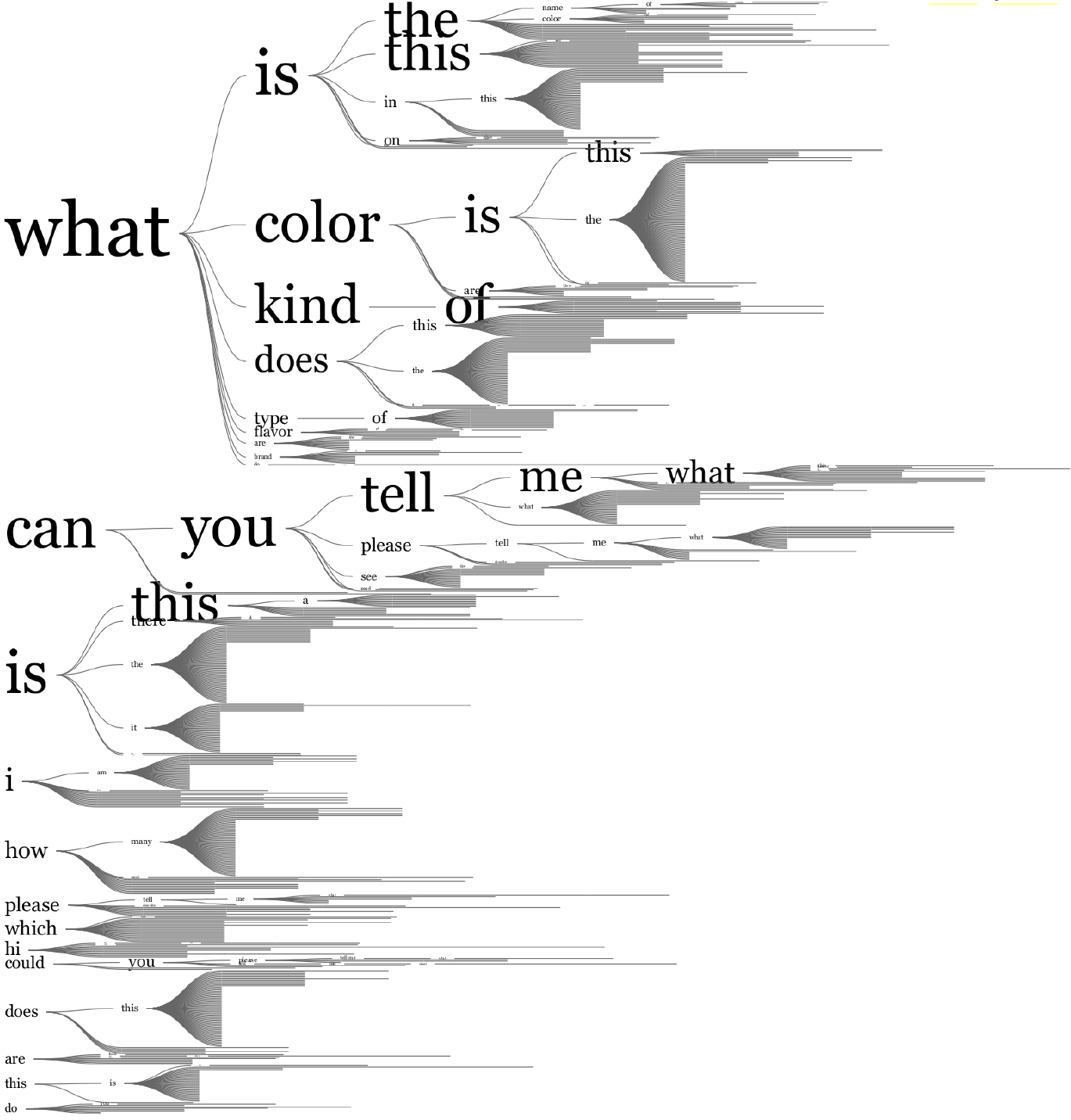}}
	\subfloat[VQA2.0]{\includegraphics[width=0.4\textwidth]{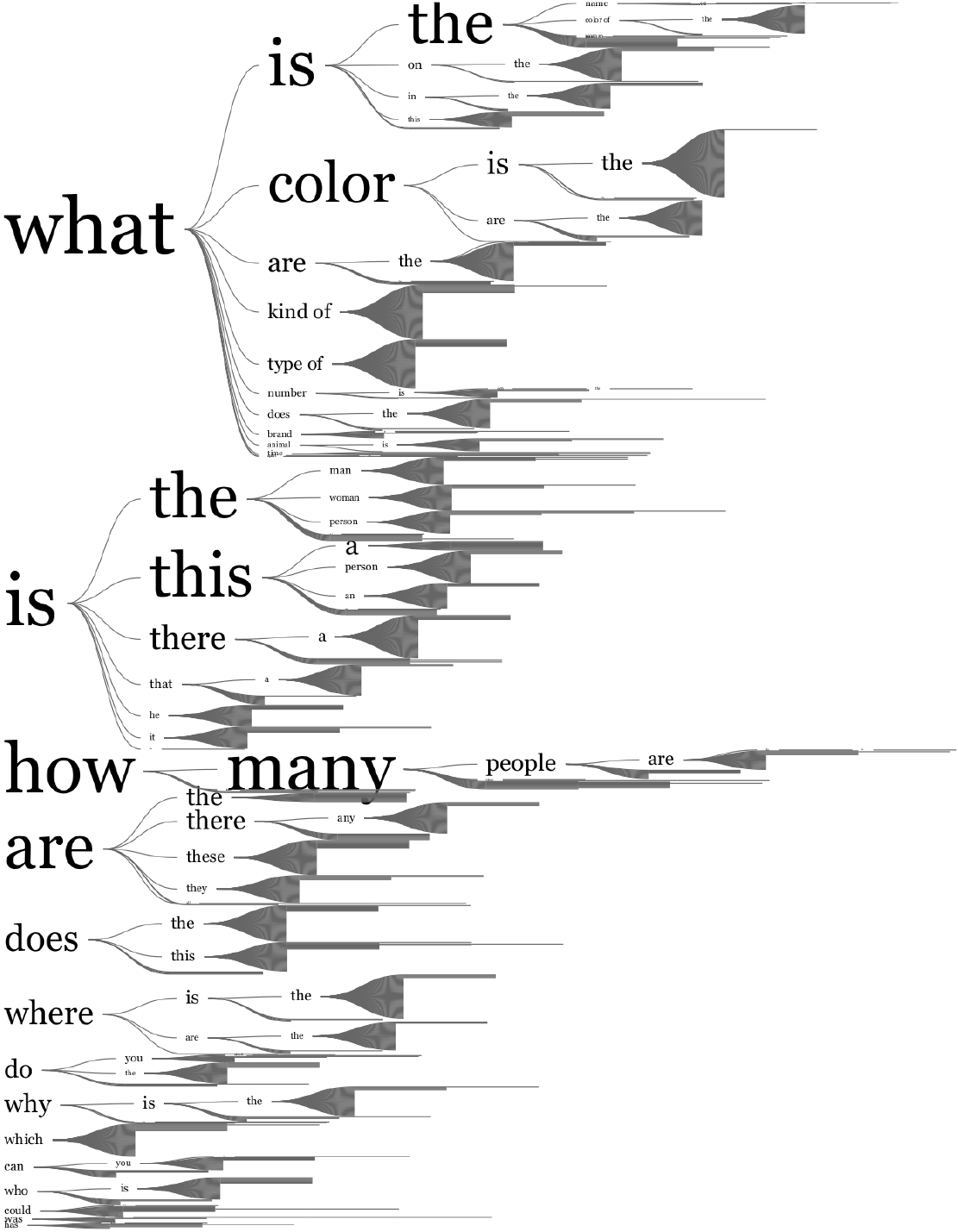}}	\quad
	\subfloat[VizWiz]{\includegraphics[width=0.4\textwidth]{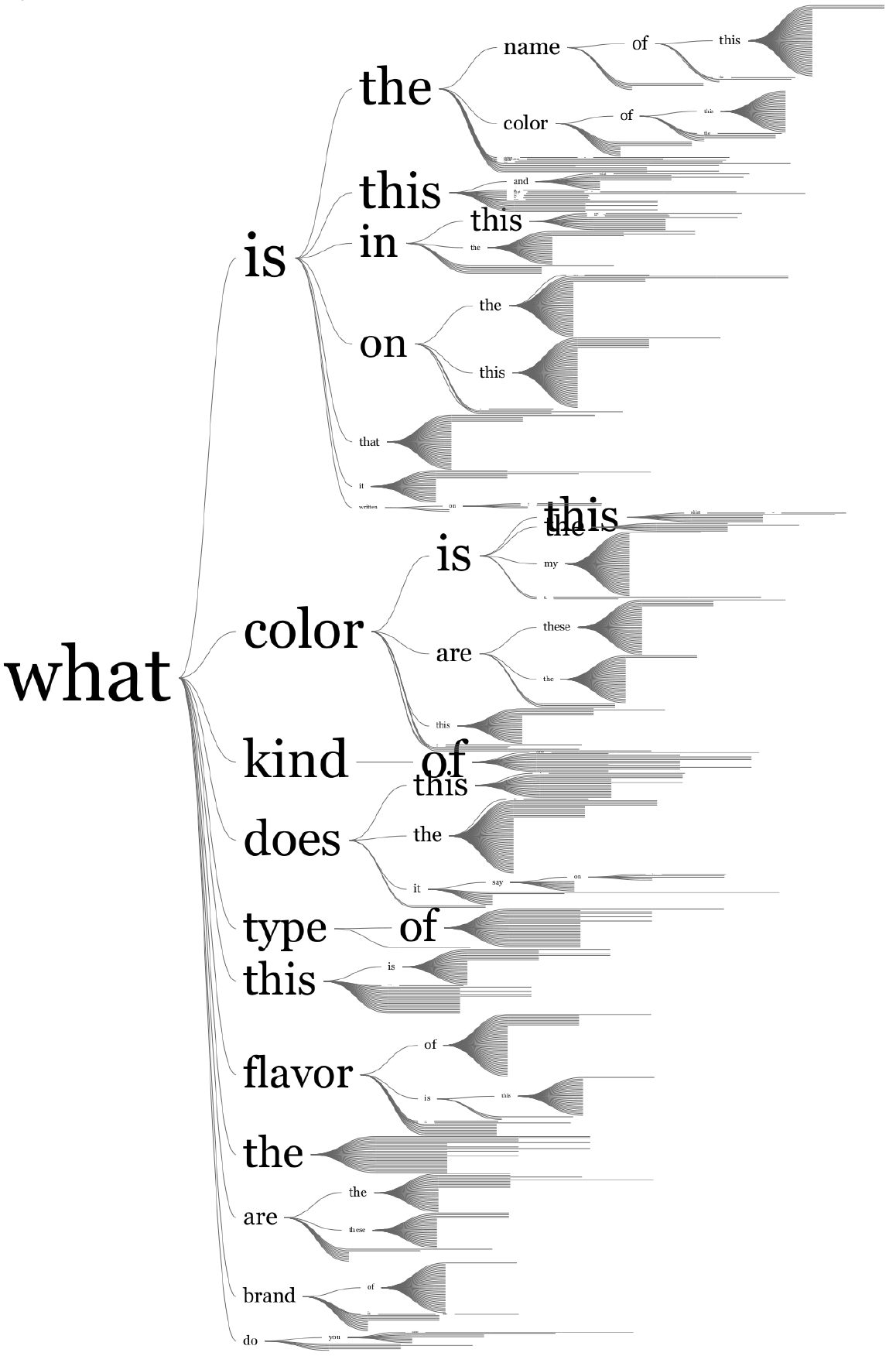}}
	\subfloat[VQA2.0]{\includegraphics[width=0.4\textwidth]{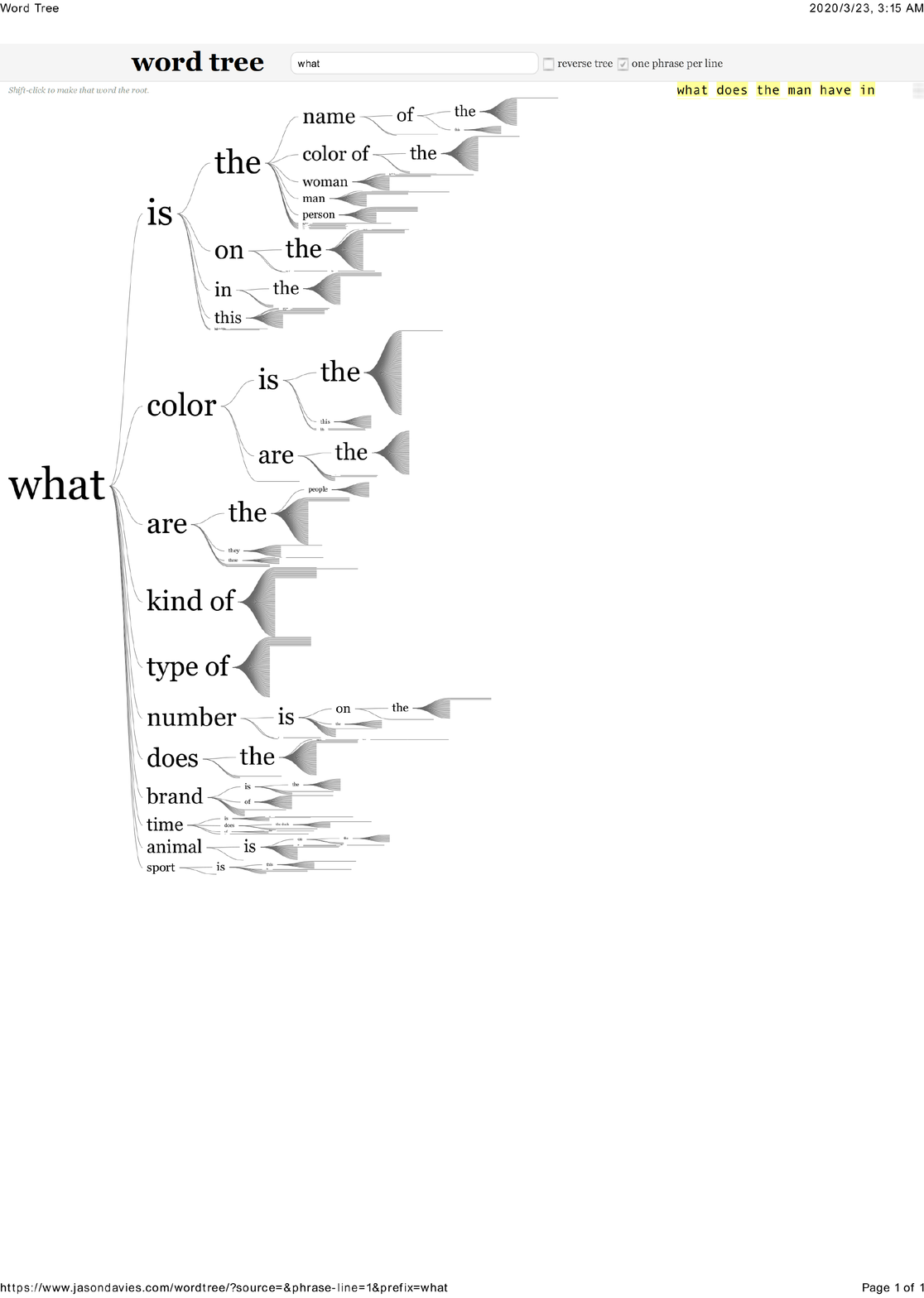}}	
	\caption{Word trees showing the most common questions requiring object recognition for both datasets (a, b) overall and (c, d) with respect to the most common first word.  The font size of the word represents the number of times the word appears.}
	\label{fig_wordTree_obj}
\end{figure}

\begin{figure}[htbp]
	\centering
	\subfloat[VizWiz]{\includegraphics[width=0.4\textwidth]{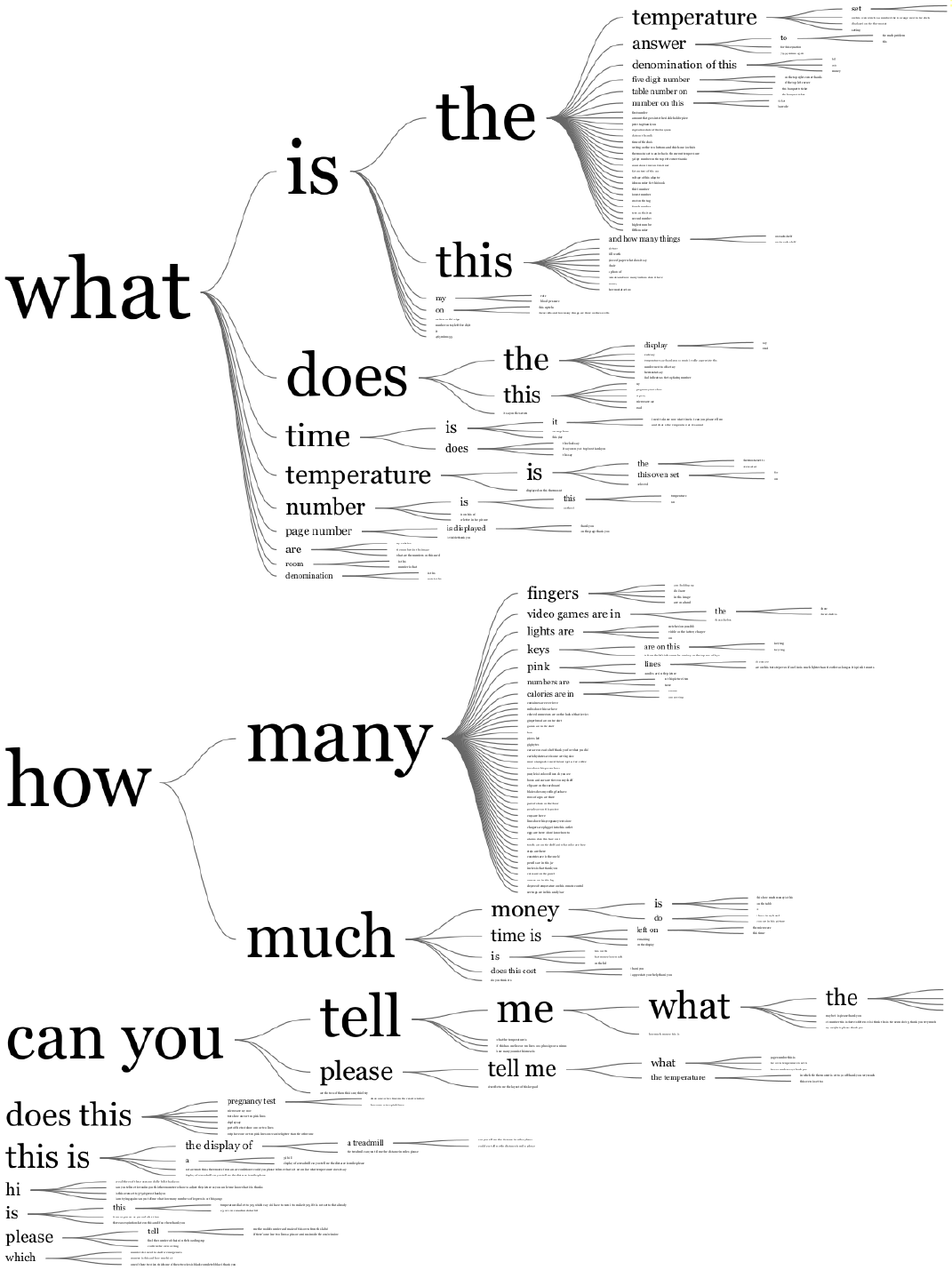}}
	\subfloat[VQA2.0]{\includegraphics[width=0.4\textwidth]{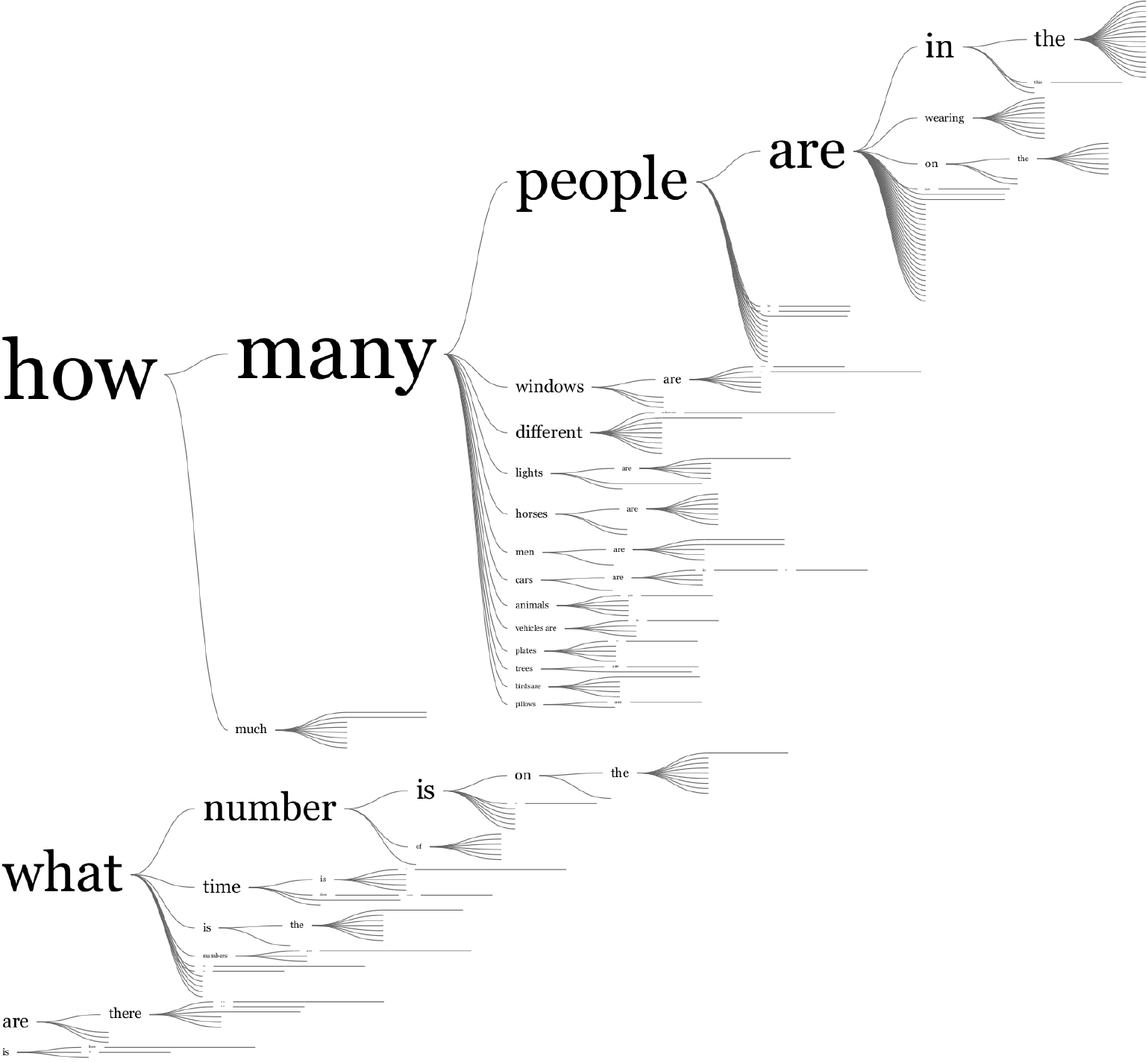}} \quad
	\subfloat[VizWiz]{\includegraphics[width=0.4\textwidth]{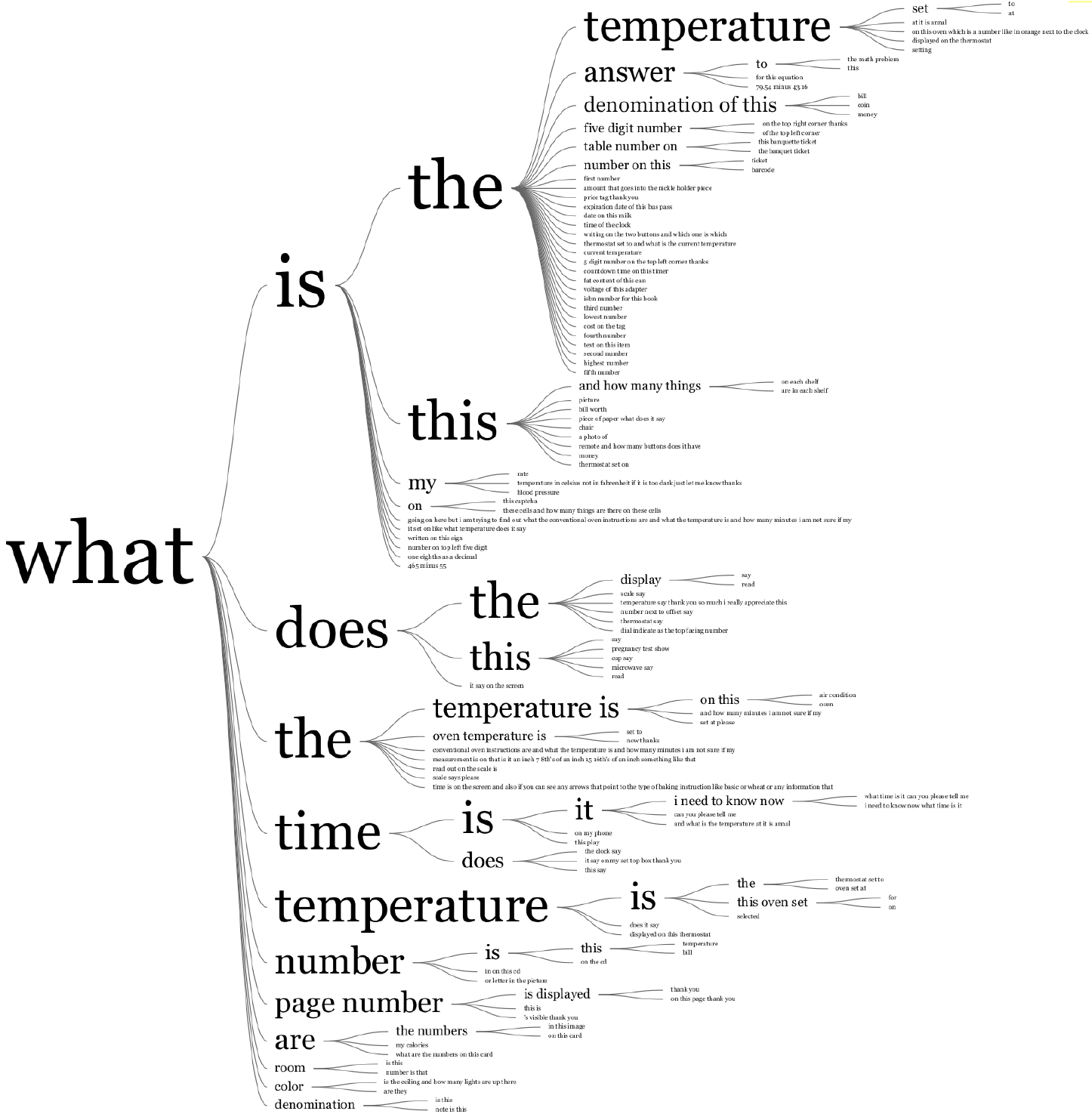}}
	\subfloat[VQA2.0]{\includegraphics[width=0.4\textwidth]{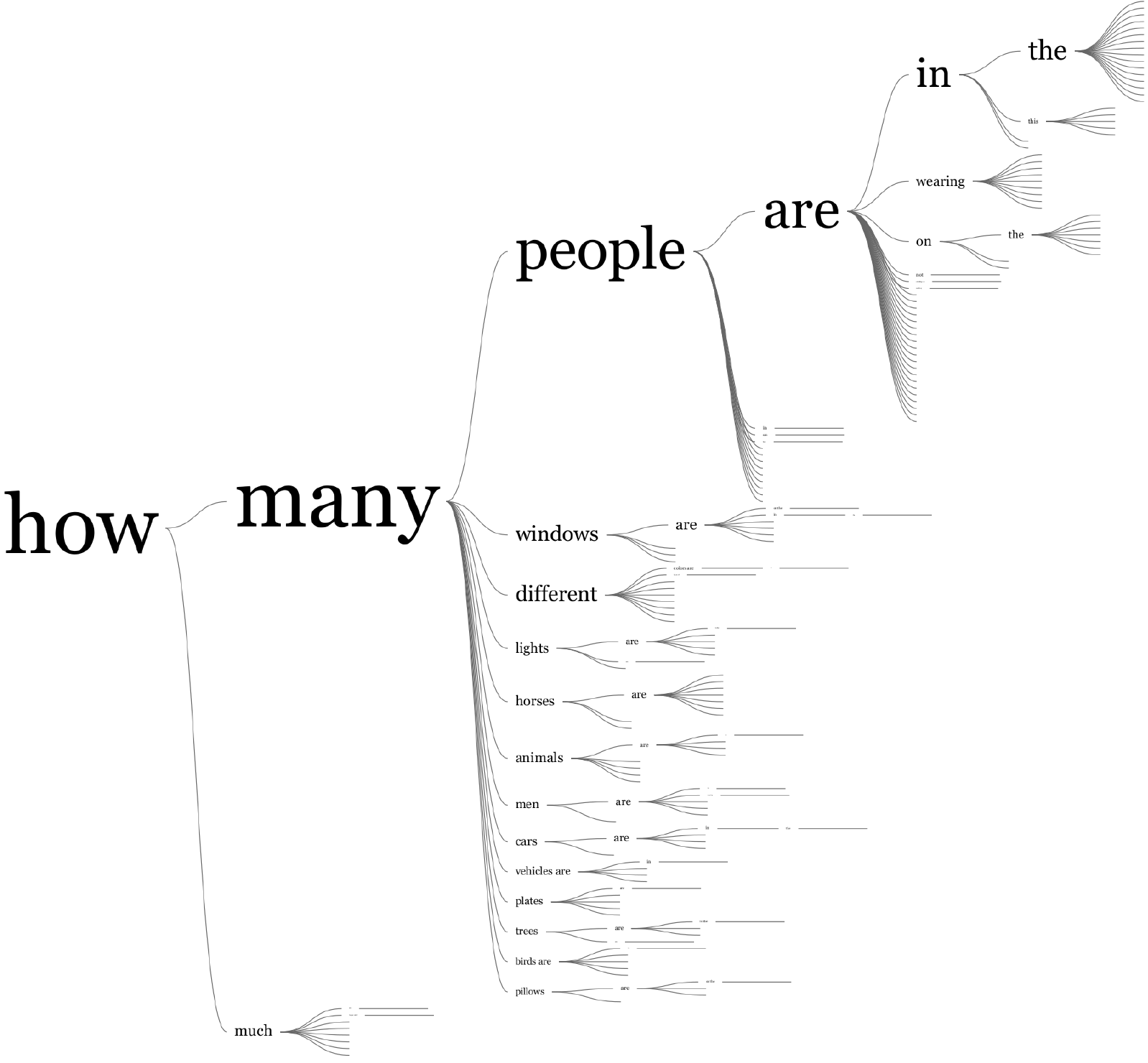}}	
	\caption{Word trees showing the most common questions requiring counting for both datasets (a, b) overall and (c, d) with respect to the most common first word.  The font size of the word represents the number of times the word appears.}
	\label{fig_wordTree_cnt}
\end{figure}

\begin{figure}[htbp]
	\centering
	\subfloat[VizWiz]{\includegraphics[width=0.4\textwidth]{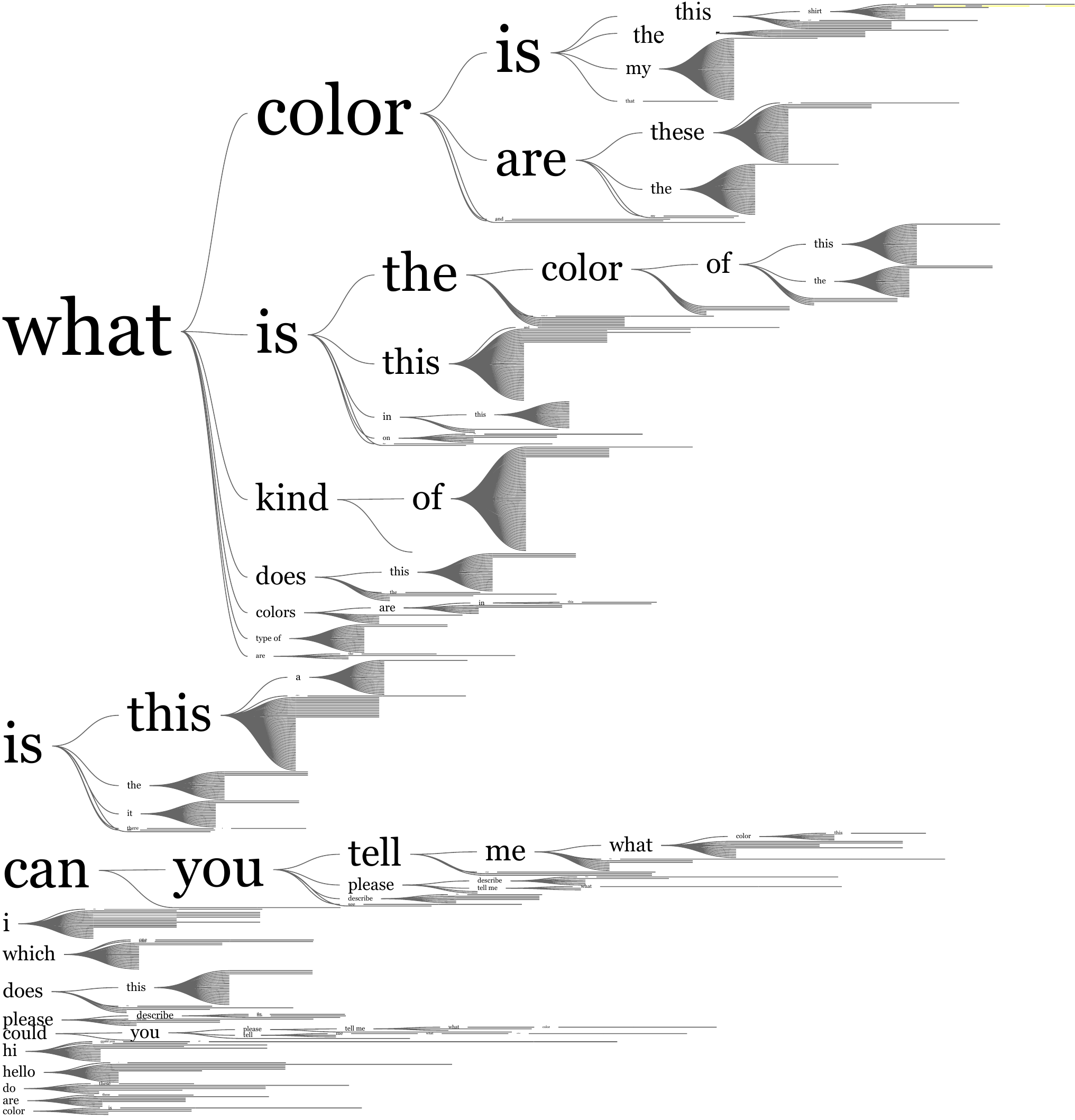}}
	\subfloat[VQA2.0]{\includegraphics[width=0.4\textwidth]{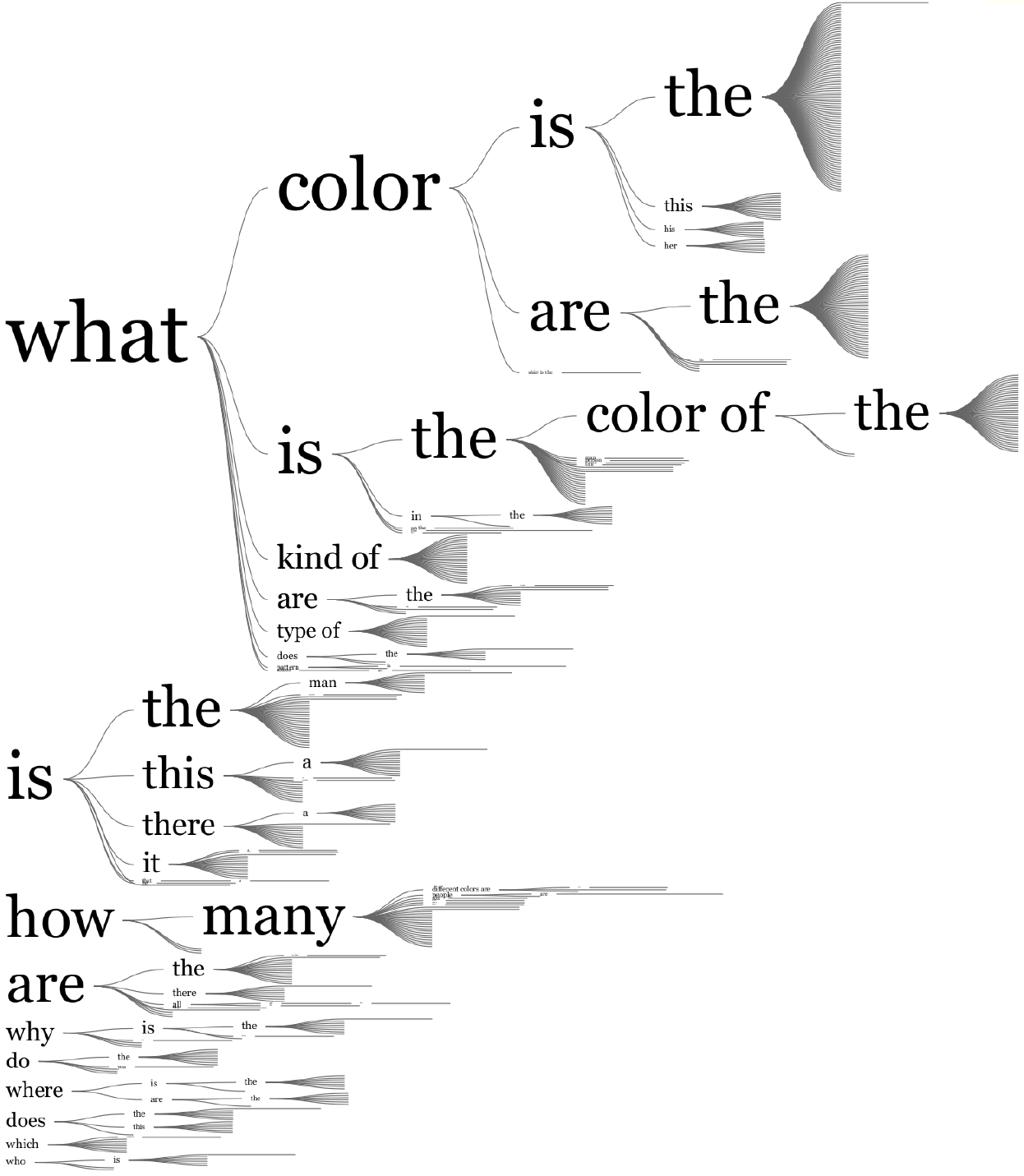}} \quad
	\subfloat[VizWiz]{\includegraphics[width=0.4\textwidth]{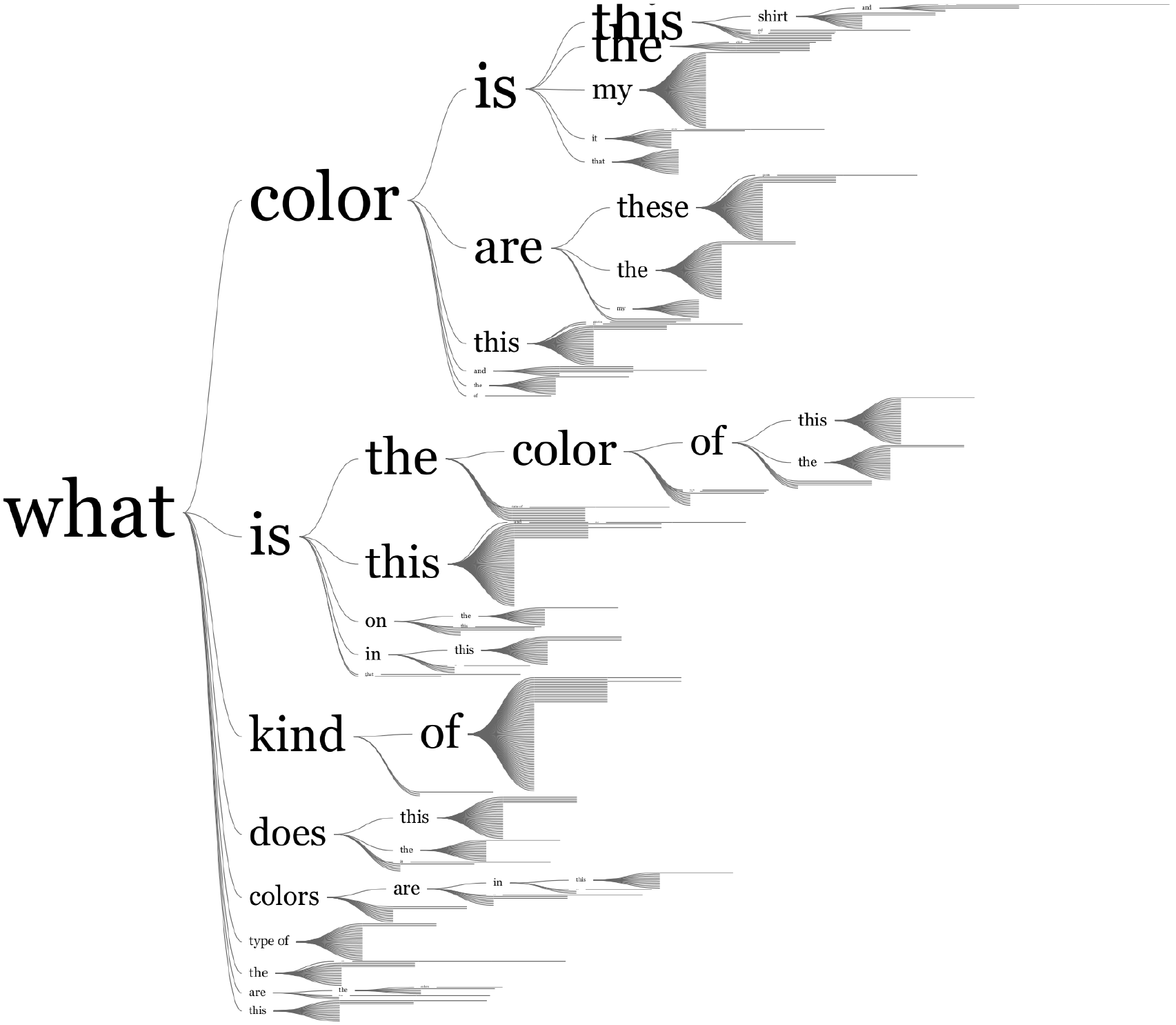}}
	\subfloat[VQA2.0]{\includegraphics[width=0.4\textwidth]{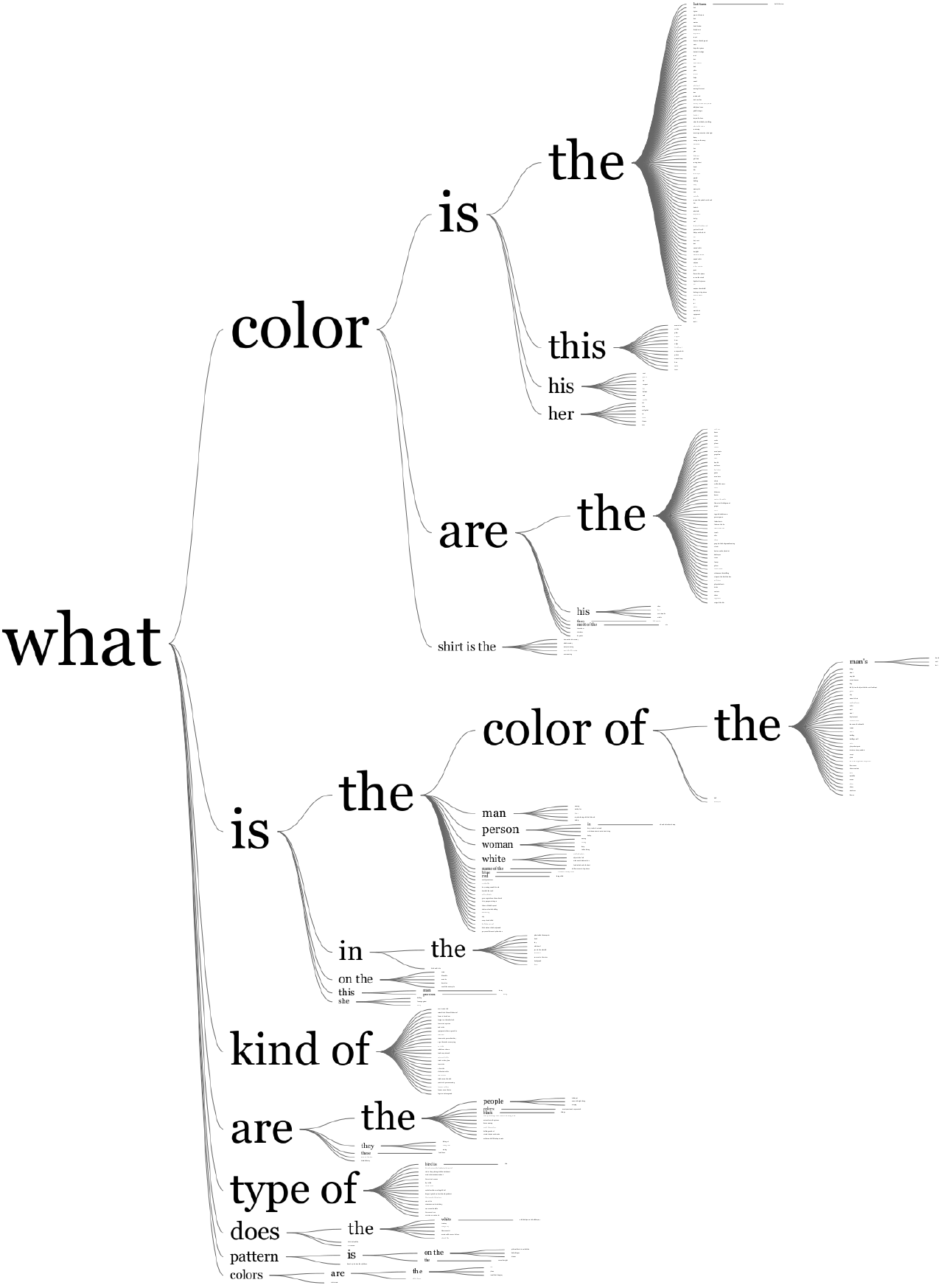}}
	\caption{Word trees showing the most common questions requiring color recognition for both datasets (a, b) overall and (c, d) with respect to the most common first word.  The font size of the word represents the number of times the word appears.}
	\label{fig_wordTree_clr}
\end{figure}

\begin{figure}[htbp]
	\centering
	\subfloat[VizWiz]{\includegraphics[width=0.4\textwidth]{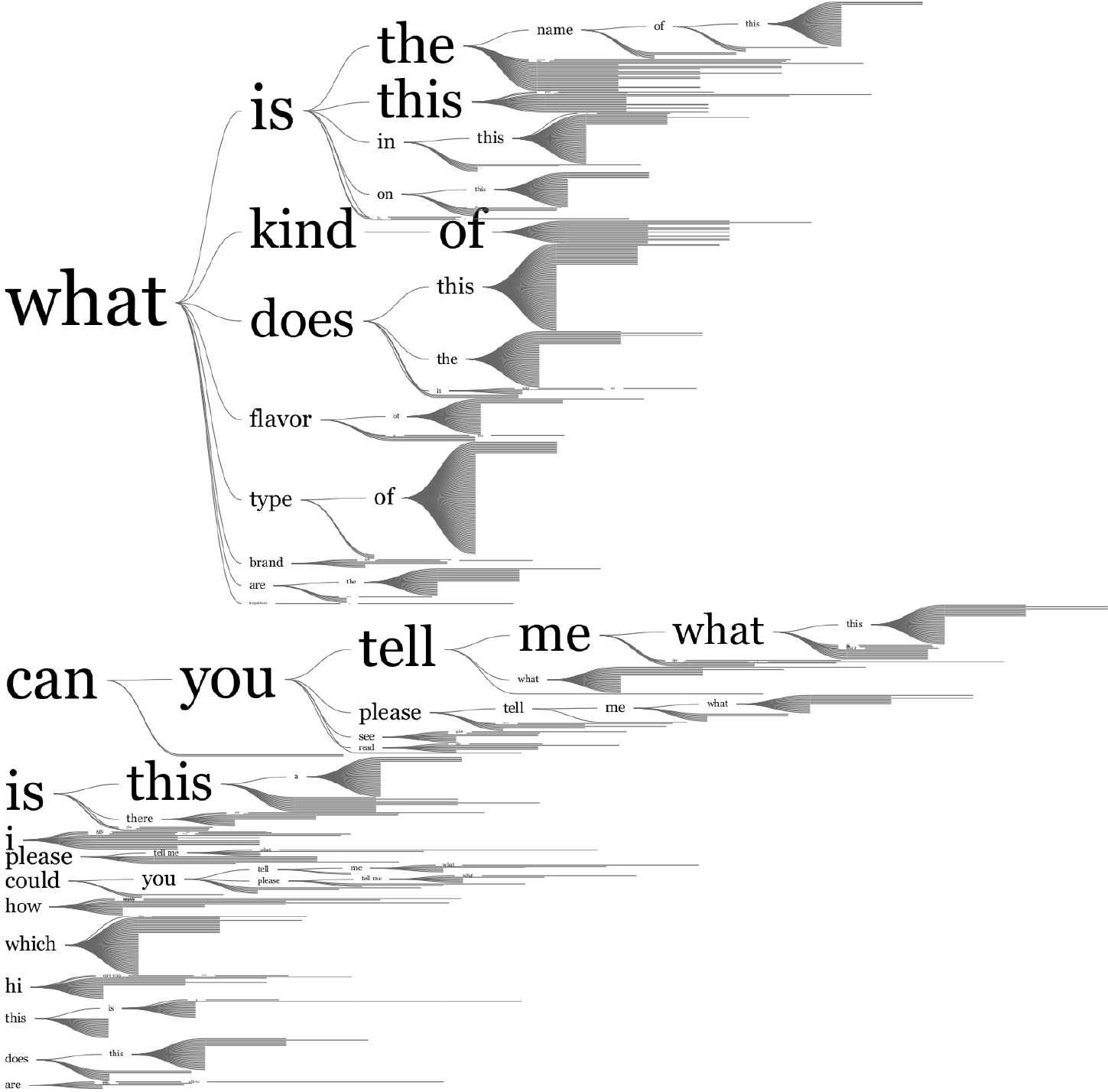}}
	\subfloat[VQA2.0]{\includegraphics[width=0.4\textwidth]{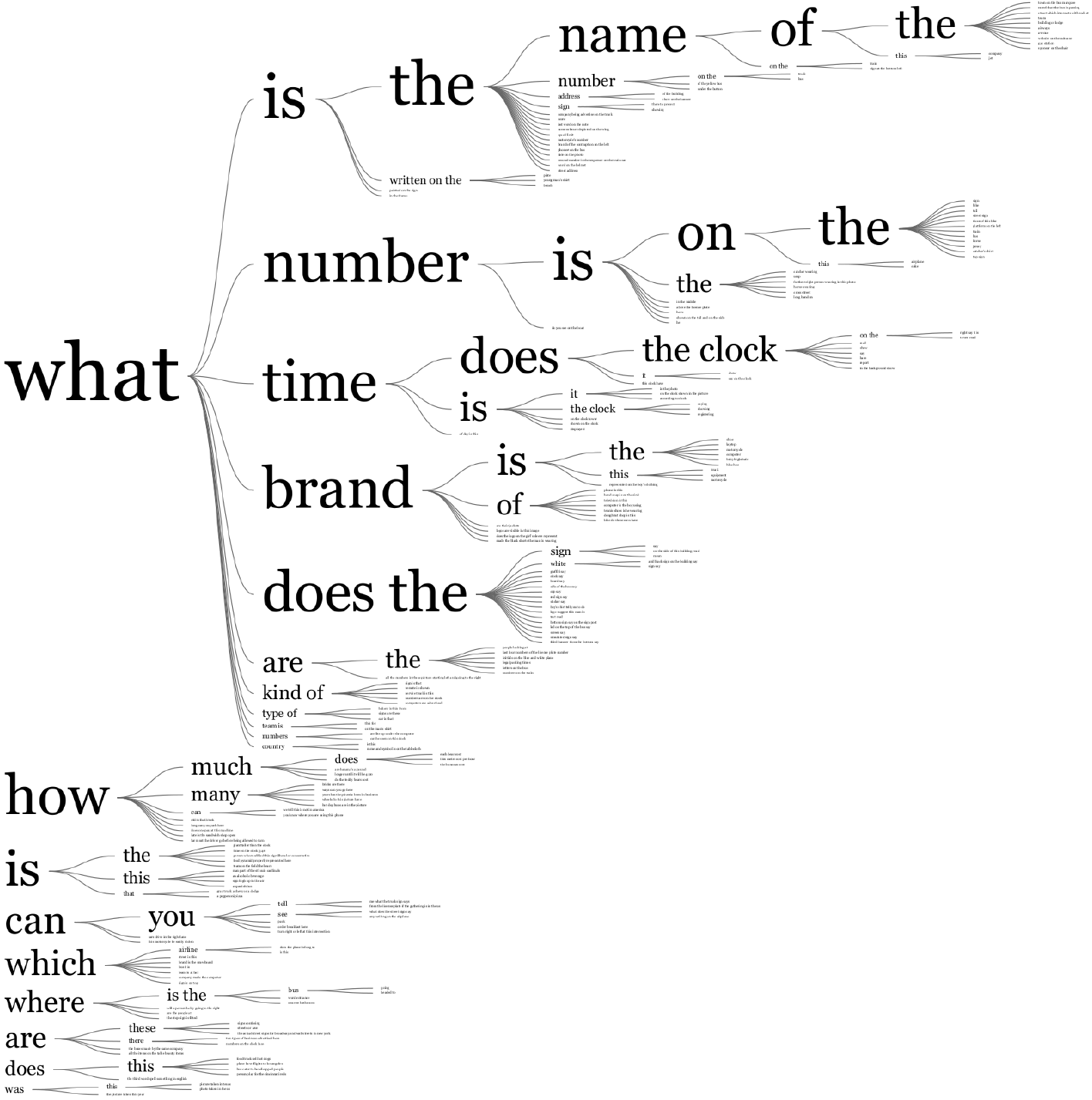}} \quad
	\subfloat[VizWiz]{\includegraphics[width=0.4\textwidth]{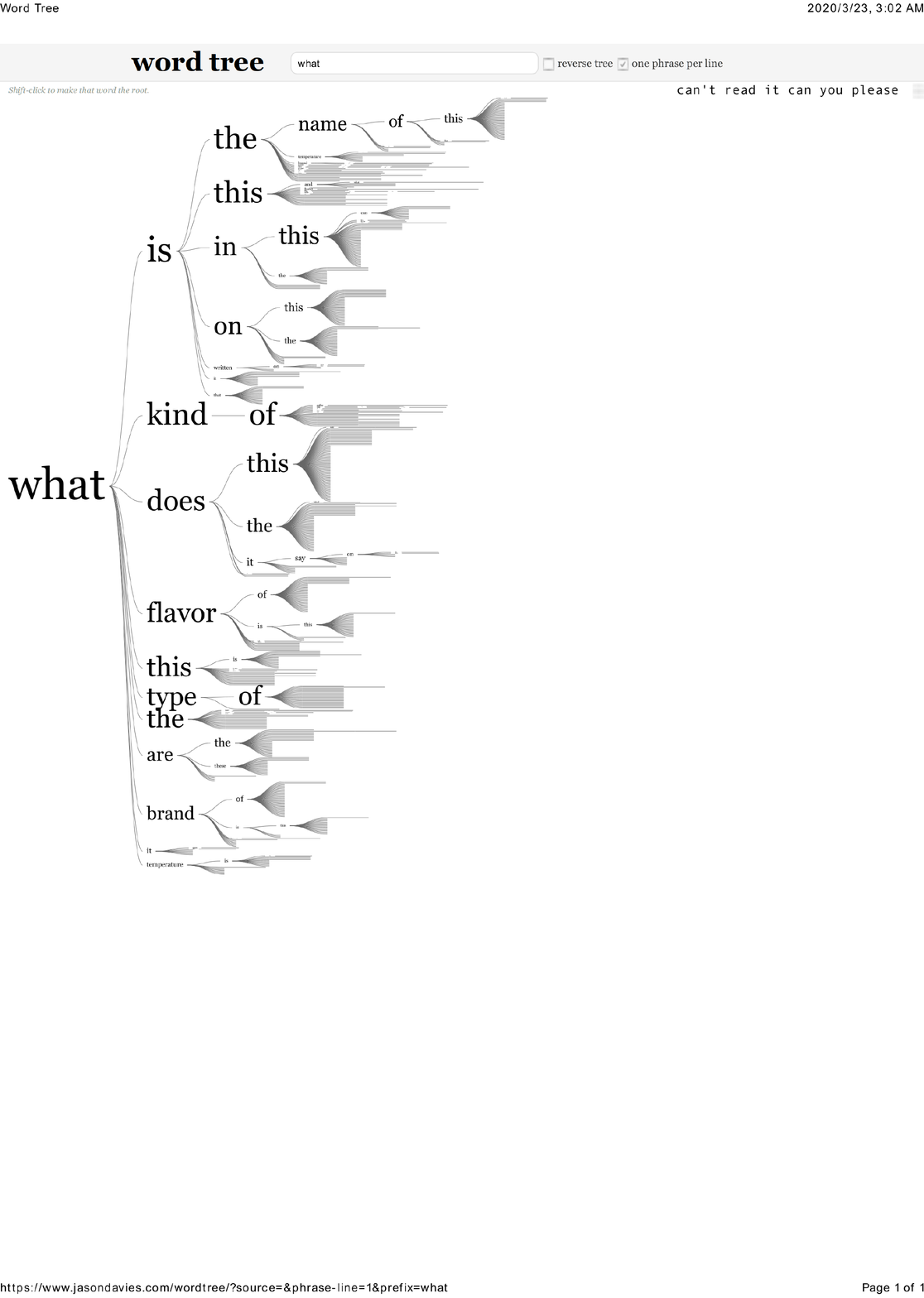}}
	\subfloat[VQA2.0]{\includegraphics[width=0.4\textwidth]{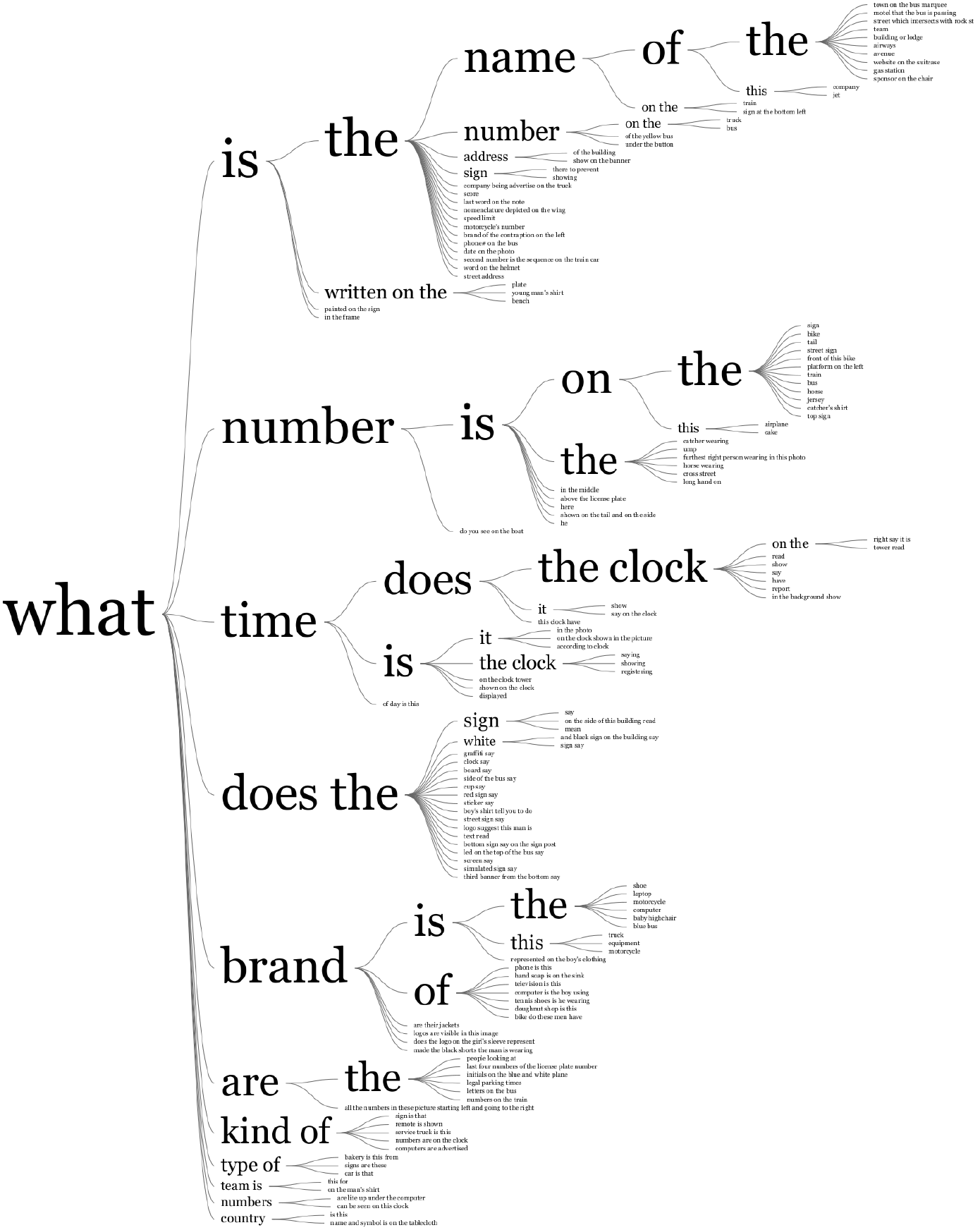}}
	\caption{Word trees showing the most common questions requiring text recognition for both datasets (a, b) overall and (c, d) with respect to the most common first word.  The font size of the word is used to represent the number of times the word appears.}
	\label{fig_wordTree_txt}
\end{figure}

To expand upon our analysis about the questions that are asked for each skill (in Figure~\ref{fig:sunburstplots} of the main paper), we provide word trees \cite{wattenberg2008word} that show the most common questions derived from the first words for each skill.  The font size of each word represents the number of times the word appears, with the size ``proportional to the square root of the frequency of the word''~\cite{wattenberg2008word}.  Results for all skills are shown in Figures~\ref{fig_wordTree_obj}, \ref{fig_wordTree_cnt}, \ref{fig_wordTree_clr}, and \ref{fig_wordTree_txt}.

\subsection*{Comparisons of the Difficulty of Different Skills for Humans}
We conducted statistical testing to expand upon our comparisons of the difficulty of the different skills for humans.  Results and corresponding p-values are shown in Tables~\ref{stat-text} and \ref{stat-diffData}. We conducted Levene's test to evaluate the equality of variances first. Then based on different situations of equality of variances, we conducted different t-tests to verify the statistical significance of the difference between the mean values of difference groups. For groups that have statistically unequal variances, we conducted Welch’s t-test between the groups; and for groups whose variances are statistically equal, we conducted Student's t-test.

\begin{table}[htbp]
\begin{tabular}{c|cccc}
\hline
                                  &                     & \textbf{\begin{tabular}[c]{@{}c@{}}Object Recognition \\ v.s.\\ Text Recognition\end{tabular}} & \textbf{\begin{tabular}[c]{@{}c@{}}Color Recognition \\ v.s.\\ Text Recognition\end{tabular}} & \textbf{\begin{tabular}[c]{@{}c@{}}Counting \\ v.s.\\ Text Recognition\end{tabular}} \\ \hline
\multirow{2}{*}{\textbf{VizWiz}}  & \textbf{Levene's test} & \begin{tabular}[c]{@{}c@{}}22.58\\ ($p < .001$)\end{tabular}                                 & \begin{tabular}[c]{@{}c@{}}186.34\\ ($p < .001$)\end{tabular}                               & \begin{tabular}[c]{@{}c@{}}9.21\\ ($p = 0.00242$)\end{tabular}                        \\ \cline{2-5} 
                                  & \textbf{t-Test}     & \begin{tabular}[c]{@{}c@{}}Welch's t = 3.25\\ ($p =0.0118 $)\end{tabular}                      & \begin{tabular}[c]{@{}c@{}}Welch's t = -5.17\\ ($p < .001$)\end{tabular}                    & \begin{tabular}[c]{@{}c@{}}Welch's t = -5.41\\ ($p < .001$)\end{tabular}           \\ \hline
\multirow{2}{*}{\textbf{VQA2.0}} & \textbf{Levene's test} & \begin{tabular}[c]{@{}c@{}}0.25\\ ($p$ = 0.62)\end{tabular}                                      & \begin{tabular}[c]{@{}c@{}}0.32\\ ($p$ = 0.57)\end{tabular}                                     & \begin{tabular}[c]{@{}c@{}}0.09\\ ($p$ = 0.76)\end{tabular}                            \\ \cline{2-5} 
                                  & \textbf{t-Test}     & \begin{tabular}[c]{@{}c@{}}Student's t = -7.87\\ ($p < .001$)\end{tabular}                   & \begin{tabular}[c]{@{}c@{}}Student's t = -5.91\\ ($p < .001$)\end{tabular}                  & \begin{tabular}[c]{@{}c@{}}Student's t = -3.60\\ ($p < .001$)\end{tabular}         \\ \hline
\end{tabular}
\caption{Statistical testing results for comparisons of human difficulty between text recognition and other skills in the same dataset. We used a significance level of $0.1\%$. The sample size is $22,226$ for VizWiz, and $5,034$ for VQA2.0.}
\label{stat-text}
\end{table}

\begin{table}[htbp]
\begin{tabular}{ccc}
\hline
                    & \multicolumn{2}{c}{\textbf{VQA2.0 vs VizWiz}}                                                                                                            \\
                    & \textbf{Object Recognition}                                                  & \textbf{Color Recognition}                                                 \\ \hline
\textbf{Levene's test} & \begin{tabular}[c]{@{}c@{}}268.24\\ ($p < .001$)\end{tabular}              & \begin{tabular}[c]{@{}c@{}}135.28\\ ($p < .001$ )\end{tabular}            \\ \hline
\textbf{t-Test}     & \begin{tabular}[c]{@{}c@{}}Welch's t = -21.50\\ ($p < .001$)\end{tabular} & \begin{tabular}[c]{@{}c@{}}Welch's t = -3.30\\ ($p < .001$)\end{tabular} \\ \hline
\end{tabular}
\caption{Statistical testing results for comparisons of human difficulty of object recognition and color recognition between the two datasets. We used a significance level of $0.1\%$. The sample size is $22,226$ for VizWiz, and $5,034$ for VQA2.0.} \begin{flushleft} \par\bigskip \noindent Received January 2020; revised June 2020; accepted July 2020. \end{flushleft}
\label{stat-diffData}
\end{table}

\end{document}